\documentclass[a4paper,11pt]{article}
\pdfoutput=1 
             
\usepackage{jheppub} 
\usepackage[T1]{fontenc}

\usepackage{xcolor}
\usepackage[utf8]{inputenc}
\usepackage{amsmath}
\usepackage{amssymb}
\usepackage{graphicx}
\usepackage{multirow}
\usepackage[normalem]{ulem}
\usepackage{comment}
\usepackage{slashed}
\usepackage{pifont}
\usepackage{array}
\usepackage{makecell}

\usepackage{tikz-feynman}
\tikzfeynmanset{compat=1.1.0}
\usetikzlibrary{arrows,automata,positioning}
\tikzset{
    vertex/.style={circle,draw, minimum size=1.5em},
    edge/.style={->,> = latex'}
}

\newcommand{\med}[1]{\langle #1\rangle}

\newcommand{\braket}[1]{\ensuremath{\left\langle #1 \right\rangle}}

\def\r {\rightarrow}

\newdimen\arrowsize
\newdimen\mylw
\pgfkeys{/my arrows/chemeq/.style={draw,thick,double distance=2pt,onearc-onearc}}
\pgfkeys{/my arrows/size/.code={\pgfsetarrowoptions{onearc}{#1}}}
\def\myalw{.4pt}
\pgfarrowsdeclare{onearc}{onearc}{%
  \mylw=\myalw
  \pgfarrowsleftextend{-\pgfgetarrowoptions{onearc}-.5\mylw}
  \pgfarrowsrightextend{0pt}
}{%
  \pgfsetdash{}{0pt}
  \mylw=\pgflinewidth
  \pgfsetlinewidth{\myalw}
  \advance\arrowsize by.5\pgflinewidth
  \pgfpathmoveto{\pgfpoint{-\pgfgetarrowoptions{onearc}}{-\pgfgetarrowoptions{onearc}-.5\mylw}}%
  \pgfpatharc{180}{90}{\pgfgetarrowoptions{onearc}}
  \pgfusepathqstroke
}

\preprint{IFIC/23-03}

\title{\boldmath  
	 Asymmetries in Extended Dark Sectors: \\A Cogenesis Scenario}

\author{Juan Herrero-Garc\'{\i}a\,,}
\author{Giacomo Landini}
\author{and Drona Vatsyayan}
\affiliation{Departamento de F\'isica Te\'orica, Universidad de Valencia and IFIC, Universidad de Valencia-CSIC,
C/ Catedr\'atico Jos\'e Beltr\'an, 2 | E-46980 Paterna, Spain}

\emailAdd{juan.herrero@ific.uv.es}
\emailAdd{giacomo.landini@ific.uv.es}
\emailAdd{drona.vatsyayan@ific.uv.es}

\abstract{The observed dark matter relic abundance may be explained by different mechanisms, such as thermal freeze-out/freeze-in, with one or more symmetric/asymmetric components. In this work we investigate the role played by asymmetries in determining the yield and nature of dark matter in non-minimal scenarios with more than one dark matter particle. In particular, we show that the energy density of a particle may come from an asymmetry, even if the particle is asymptotically symmetric by nature. To illustrate the different effects of asymmetries, we adopt a model with two dark matter components. We embed it in a multi-component cogenesis scenario that is also able to reproduce neutrino masses and the baryon asymmetry. In some cases, the model predicts an interesting monochromatic neutrino line that may be searched for at neutrino telescopes.
}


\makeatletter
\gdef\@fpheader{}
\makeatother

\begin{document}
\maketitle
\flushbottom

\section{Introduction}{\label{sec:intro}}

The nature of Dark Matter (DM), that makes up roughly a quarter of the energy density of our universe, along with the origin of neutrino masses and the baryon asymmetry (BAU), are among the most important open problems that the Standard Model (SM) fails to explain, with overwhelming experimental evidence. Several extensions of the SM have been proposed, where either a single particle or several stable particles, i.e., multi-component DM \cite{BasiBeneito:2022qxd,Cao:2007fy,Zurek:2008qg,Belanger:2011ww,Liu:2011aa,Arcadi:2016kmk,Bhattacharya:2016ysw,Bernal:2018aon,Borah:2019aeq}, make up the observed DM relic abundance, $\Omega_{\rm DM}h^2 \sim 0.1$ \cite{Planck:2018vyg}. The most popular mechanisms to reproduce this relic abundance include thermal freeze-out (FO) of weakly interacting massive particles (WIMPs) \cite{Arcadi:2017kky, Roszkowski:2017nbc,Buttazzo:2019iwr,Landini:2020daq,Coito:2021fgo,Coito:2022kif} and freeze-in (FI) of feebly interacting massive particles (FIMPs) \cite{Hall:2009bx, Bernal:2017kxu,Gross:2020zam}. 

WIMPs are initially in thermal equilibrium with the SM and undergo annihilations until they freeze-out when the annihilation rate drops below the Hubble expansion rate; therefore, the DM abundance is inversely proportional to the annihilation rate. On the other hand, FIMPs have a negligible initial abundance and are produced mainly by tiny interactions with particles in the thermal bath so that they never thermalise, and they freeze-in once the mother particle decouples from the bath; thus, the DM abundance is directly proportional to the production rate. Depending on the interactions, a further contribution may come from late decays (LD) of the mother particle, the size of which is model-dependent. 

Most of these models assume that the new states are symmetric in nature, i.e., the abundance of the DM particle is the same as that of the antiparticle. However, there exist a wide variety of models that propose that the DM abundance is rather set by an initial asymmetry in the dark sector, in analogy to the visible sector (where $\eta_B = 0.88 \times 10^{-11}$), motivated by the closeness of baryonic and DM energy densities, $\rho_{\rm DM} \sim 5 \rho_B$ \cite{Kaplan:2009ag}. The asymmetry can be first generated in the visible sector and then transferred to dark sector (or vice versa) \cite{Feng:2012jn,Blennow:2010qp,Hall:2021zsk}, or an asymmetry can be generated simultaneously in both the sectors (\textit{cogenesis}). Such asymmetric dark matter (ADM) models (see Refs.~\cite{Petraki:2013wwa,Zurek:2013wia} for a review) aim to explain the observed baryon asymmetry and DM abundance in a common framework. 

An intermediate scenario between the two extremes involves the asymmetric freeze-out of a species, where the DM particle and its antiparticle freeze-out with different number densities, depending on the initial asymmetry, and the dark matter is partially asymmetric \cite{Graesser:2011wi}. Hence, a further distinction can be made regarding the nature of DM, whether it is symmetric, asymmetric or partially asymmetric. Therefore, the presence of an asymmetry can have significant implications for the mechanisms discussed above in reproducing the DM abundance. 

The possibility that asymmetric single-component WIMP DM  is produced in a cogenesis scenario has been studied in Ref.~\cite{Cui2020}. Another model has been presented in Ref.~\cite{Falkowski:2011xh}, with the possibility to restore the symmetric nature of DM through late decays of an extra particle. 
In Ref.~\cite{Borah2019} a model of symmetric multi-component DM, which combines freeze-out and freeze-in, accompanied by the generation of the baryon asymmetry, has been proposed. Finally, Refs.~\cite{Shuve:2020evk,Goudelis:2021qla,Chand:2022vrf} study models linking the freeze-in production of dark matter with baryogenesis; however, the dark matter in  these models is symmetric.

In this work, we aim to generalize this picture, starting from the concrete cogenesis scenario realized in Ref.~\cite{Falkowski:2011xh}, by considering an extended dark sector in which we can realize multicomponent DM, so that one DM component, or even both, can be asymmetric. Furthermore, we combine freeze-out and freeze-in production, including the possibility of asymmetric freeze-in. The idea of asymmetric freeze-in was proposed in Ref.~\cite{Hall:2010jx}, mainly focusing on simultaneously generating equal and opposite asymmetries in the visible and dark sectors, which are separately in thermal equilibrium at different temperatures. Unlike the case of asymmetric freeze-out, it was shown that this case is not so straightforward and it requires the presence of a richer dark sector, so the dark matter particle is able to thermalise with the dark states even if it is feebly coupled to the SM particles. In Ref.~\cite{Hook:2011tk}, the unitarity and CPT constraints for transferring the baryon number to the dark sector via freeze-in were discussed. In Ref.~\cite{Unwin:2014poa}, it was shown that dark sector asymmetries produced by asymmetric freeze-in can be sizeable and comparable to the baryon asymmetry if the mediator between the two sectors is out-of-equilibrium. However, these works mainly focus on the generation of asymmetry but do not study the details of the annihilation of the symmetric component, which is a necessary ingredient for any ADM mechanism.
   
To this end, we propose a model in which the dark sector is connected to the visible sector via a mediator particle. The asymmetries in both sectors are generated via cogenesis, which yields a relation between neutrino masses, the baryon asymmetry and the DM relic abundance. In addition to the generation of asymmetries, we also discuss the annihilation of the symmetric component in a complete model with all the necessary ingredients. We propose a feebly interacting particle that can be an asymmetric DM candidate with no sizeable couplings with neither the SM particles nor the dark sector ones. We then investigate the role of the dark sector asymmetry in determining the relic abundance of one/several particles. For example, depending on the model parameters, the DM may either be single-component or multi-component, with either all symmetric or asymmetric components or a mixture of both. Similarly, the production mechanisms may be freeze-out, freeze-in or some via freeze-out and the others via freeze-in. Moreover, we show that in certain scenarios the particle abundance may be set by an asymmetry even if its nature is symmetric. We find that some of them predict the existence of an observable neutrino line.

The paper is structured as follows. In Section~\ref{sec:fram}, we elaborate on the framework of generating an asymmetry and the possibility of asymmetric freeze-in. We discuss the complete model, which also generates neutrino masses and the baryon asymmetry, in Section~\ref{sec:model}. The different DM candidates are discussed in Section~\ref{sec:darkmatter}. The contribution to the DM relic abundance is studied in Section~\ref{sec:dmrelic}. In section~\ref{sec:pheno}, we discuss the main phenomenological signatures, with special emphasis on the prediction of a monochromatic neutrino line. In Section~\ref{sec:ISS}, we discuss a low scale variant of the model (an inverse seesaw). Finally, we give our conclusions in Section~\ref{sec:conclusions}. We show further details in some appendices.

\section{General framework}\label{sec:fram}

In order to generate a dark sector asymmetry, we focus on the cogenesis of both DM and baryon asymmetries and adopt the two-sector thermal leptogenesis mechanism of Ref.~\cite{Falkowski:2011xh}. In this case, the asymmetries are produced from out-of-equilibrium CP-violating decays of right handed neutrinos (RHNs), which are well-motivated to reproduce neutrino masses. This requires the leptons and some of the dark sector particles to be charged under a lepton symmetry that is broken by Majorana masses of the RHNs, thus satisfying all the conditions for the dynamical generation of an asymmetry \cite{Sakharov:1967dj}. The RHNs are initially in thermal equilibrium, and once the temperature drops below the mass of the lightest RHN, $M_{N_1}$, the washout and other interactions leading to transfer of asymmetries between the two sectors become inefficient. Subsequently, the asymmetries get frozen in the two sectors. The leptonic asymmetry is partially converted into the baryonic one via sphaleron processes. The dark asymmetry ($\eta_i \equiv Y_i^+-Y_i^-$, where we introduce the yield as the number density upon entropy density, $Y_i = n_i/s$) is carried by a dark fermion (we denote it by $\chi$) and once its symmetric component gets annihilated away, the asymmetric one sets the relic abundance, $\Omega_{\rm DM} \propto \eta_\chi m_\chi$. This framework, therefore, connects neutrino mass generation with the baryon asymmetry and the DM relic abundance. See Refs.~\cite{Cosme:2005sb,Bhattacharya:2021jli,Falkowski:2017uya,Datta:2021elq,Biswas:2018sib,An:2009vq,Chianese:2019epo,Chun:2011cc} for other extensions of seesaw framework that address the three issues under the same umbrella.

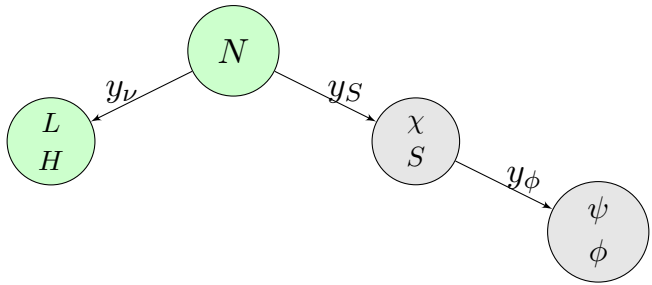
\begin{figure}[!htb]
    \centering
     \begin{tikzpicture}[scale=1.2, transform shape,every text node part/.style={align=center}]
        \node[draw,circle,fill=green!20!white,minimum size=1cm,scale=1] (P0) at (0,0) {$N$};
        \node[draw,circle,fill=green!20!white,minimum size=1.2cm,scale=0.8] (P1) at (-2,-1) {$L$\\$H$};
        \node[draw,circle,fill=gray!20!white,minimum size=1.2cm,scale=0.8] (P2) at (2,-1) {$\chi$\\$S$};       
        \node[draw,circle,fill=gray!20!white,minimum size=1cm,scale=0.9] (P3) at (4,-2) {$\psi$\\$\phi$};
        \draw[edge] (P0) -- (P1) node[pos=.4, left] {$y_\nu$};
        \draw[edge] (P0) -- (P2) node[pos=.4, right] {$y_{S}$};
        \draw[edge] (P2) -- (P3)node[pos=.4, right] {$y_{\phi}$};
    \end{tikzpicture}
    \caption{Schematic framework of cogenesis and DM production in the model at $T<M_{N_1}$ via the indicated Yukawa interactions $y_i$. Here, $L$ and $H$ are the SM lepton and Higgs doublet, respectively, whereas $S$ and $\phi$ are complex scalars belonging to the dark sector. 
}
    \label{fig:cogenesis}
\end{figure}

\begin{figure}[!htb]
    \centering
    \begin{tikzpicture}[scale=1.0, transform shape,every text node part/.style={align=center}]
    \node (A) at (-5.5,0) {Case I:};
        \node[draw,circle,fill=orange!20!white,minimum size=1.5cm,scale=0.8] (P0) at (0,0) {$\chi,\,S$\\$\phi$};
        \node[draw,circle,fill=green!20!white,minimum size=1.5cm,scale=0.8] (P1) at (-4,0) {SM\\+$N_R$}; 
        \node[draw,circle,fill=gray!40!white,minimum size=1.5cm,scale=0.8] (P2) at (4,0) {$\psi$};                 
	    \path[/my arrows/chemeq,/my arrows/size=4pt] (P0) -- (P1)node[pos=0.5,above] {Equilibrium};
        \path[/my arrows/chemeq,/my arrows/size=4pt] (P0) -- (P2)node[pos=0.5,above] {Equilibrium};
     \end{tikzpicture}
         
     \begin{tikzpicture}[scale=1.0, transform shape,every text node part/.style={align=center}]
     \node (A) at (-5.5,0) {Case II:};
        \node[draw,circle,fill=orange!20!white,minimum size=1.5cm,scale=0.8] (P0) at (0,0) {$\chi,\,S$\\$\phi$};
        \node[draw,circle,fill=green!20!white,minimum size=1.5cm,scale=0.8] (P1) at (-4,0) {SM\\+$N_R$}; 
        \node[draw,circle,fill=gray!40!white,minimum size=1.5cm,scale=0.8] (P2) at (4,0) {$\psi$};                 
	    \draw[edge] (P1) -- (P0)node[pos=0.5,above] {Freeze-in};
        \draw[edge] (P0) -- (P2)node[pos=0.5,above] {Freeze-in};
     \end{tikzpicture} 
     
     \begin{tikzpicture}[scale=1.0, transform shape,every text node part/.style={align=center}]
     \node (A) at (-5.5,0) {Case III:};
        \node[draw,circle,fill=orange!20!white,minimum size=1.5cm,scale=0.8] (P0) at (0,0) {$\chi,\,S$\\$\phi$};
        \node[draw,circle,fill=green!20!white,minimum size=1.5cm,scale=0.8] (P1) at (-4,0) {SM\\+$N_R$}; 
        \node[draw,circle,fill=gray!40!white,minimum size=1.5cm,scale=0.8] (P2) at (4,0) {$\psi$};                 
	    \path[/my arrows/chemeq,/my arrows/size=4pt] (P0) -- (P2)node[pos=0.5,above] {Equilibrium};
        \draw[edge] (P1) -- (P0)node[pos=0.5,above] {Freeze-in};
     \end{tikzpicture} 
     
      \begin{tikzpicture}[scale=1.0, transform shape,every text node part/.style={align=center}]
     \node (A) at (-5.5,0) {Case IV:};
        \node[draw,circle,fill=orange!20!white,minimum size=1.5cm,scale=0.8] (P0) at (0,0) {$\chi,\,S$\\$\phi$};
        \node[draw,circle,fill=green!20!white,minimum size=1.5cm,scale=0.8] (P1) at (-4,0) {SM\\+$N_R$}; 
        \node[draw,circle,fill=gray!40!white,minimum size=1.5cm,scale=0.8] (P2) at (4,0) {$\psi$};                 
	    \path[/my arrows/chemeq,/my arrows/size=4pt] (P0) -- (P1)node[pos=0.5,above] {Equilibrium};
        \draw[edge] (P0) -- (P2)node[pos=0.5,above] {Freeze-in};
     \end{tikzpicture} 
       
     \caption{The four scenarios for entropy transfer between the SM sector and the dark sectors formed by $\chi, S$ and $\psi, \phi$. Similar figure in Ref.~\cite{Li:2022bpp}.}
    \label{fig:ent}
\end{figure}
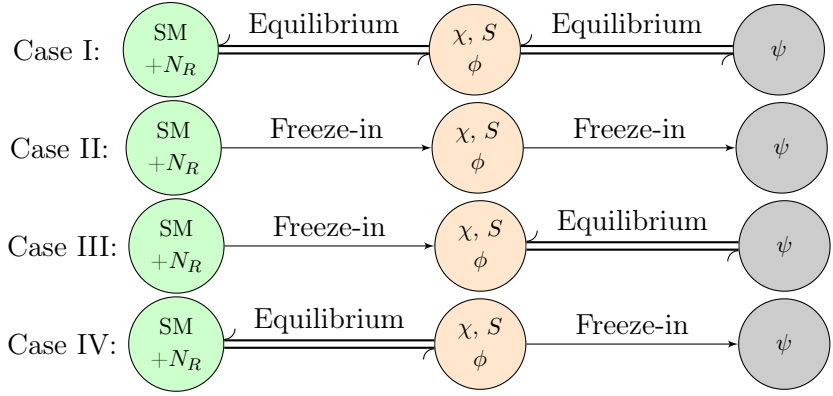

In our work, we go a step further and enlarge the dark sector so that the different possibilities discussed in the introduction are feasible. For this purpose, the particle $\chi$ in our set-up is not the dark matter but rather decays to another stable fermion (say $\psi$), which may constitute all or part of the DM abundance. The schematic framework is shown in Fig.~\ref{fig:cogenesis}. Therefore, the asymmetry in $\chi$ can be transferred to $\psi$ via its decays. 
Indeed, the set-up offers a richer phenomenology as it is now possible to accommodate multi-component DM (in this case, it could be $\psi$ and $S$), as well as different dynamics thanks to the presence and size of the different interactions. In principle, we can have four cases of equilibration between different sectors, as illustrated also in Ref.~\cite{Li:2022bpp}. In Fig.~\ref{fig:ent}, we show  the four cases of entropy transfer between the various sectors: the SM + $N_R$ (green), $\chi,S,\phi$ (orange) and $\psi$ (gray).\footnote{In the scenarios considered below, $\phi$, even if it comes from $\chi$ decays (see Fig.~\ref{fig:cogenesis}), is included in the middle blob.}

The first case involves equilibration among all sectors. It is the multi-component freeze-out scenario that has been widely considered (see for instance Ref.~\cite{BasiBeneito:2022qxd}), and we will not consider it here in the following. Cases II and III correspond to scenarios where $\chi$ is not in equilibrium with the SM + $N_R$ sector, which implies that $y_S \ll 10^{-7}$. In the considered framework where the asymmetries are produced from the decays of heavy right-handed neutrinos, such a small Yukawa coupling would then be unable to generate a dark asymmetry comparable to the visible one, and therefore DM would have to be symmetric. 
Therefore, we are interested in the last case (Case IV), where: 
\begin{itemize}
\item[\emph{i)}] the orange sector is in equilibrium with the SM + $N_R$, 
\item[\emph{ii)}] the dark sector asymmetry may be generated via co-genesis, of size comparable to the baryonic one, and
\item[\emph{iii)}] the $\psi$ sector is not in equilibrium and is produced via freeze-in. 
\end{itemize}

Whether the abundance produced by freeze-in is asymmetric depends on the value of Yukawa $y_\phi$, because the $\psi$ population can be symmetric even if the mother particle $\chi$ carries an asymmetry. This can be understood as follows. Let us define $x_i\equiv m_i/T$, for a species of mass $m_i$, where $T$ is the temperature. The asymmetry freezes out at $x_i \sim 20$, whereas the freeze-in from early decays takes place around the mass of the mother particle, i.e., $x_i \sim 1$, when both the mother particle and its antiparticle are in equilibrium. So, if the production from early decays is greater than the asymmetric yield after freeze-out, i.e., $Y_\psi =Y_{\bar{\psi}} > Y_\chi \approx \eta_D$, then the daughter particle will be symmetric in nature, because the production from late-decays (that take place at $x_i \gg 1$) will be sub-dominant and negligible. However, if the production from early decays is smaller than the asymmetry, $Y_{\rm \psi} < \eta_D$, then the late decays that are active much later after the asymmetry has frozen-in will produce more $\psi$ than $\bar{\psi}$, hence generating an asymmetry in $\psi$. Realizing this last scenario therefore yields an example of asymmetric freeze-in, which up to our knowledge has not been studied in the literature. 

In Ref.~\cite{Kitano:2004sv}, the authors discussed the idea of having an asymmetry in a messenger particle with a long enough lifetime so that it decays to the dark matter particle once its symmetric component is annihilated away. In that case, however, the non-thermal decay to dark matter was suppressed by higher-dimensional operators, which require new physics at some higher energy scale. On the other hand, in our scenario the decays are suppressed due to the feeble nature of the interaction, whose strength controls whether the contributions come from either early decays, late decays or both. This in turn determines the nature of $\psi$ population.\footnote{In Ref.~\cite{Garny:2018ali}, the contributions from early and late decays have been compared for a symmetric DM model.} Therefore, our renormalizable model does not require any extra new physics.

As we will see, in certain scenarios, the late decays of an asymmetric particle may populate the symmetric component of a species. Hence, the late decays play an important role in determining the  final nature of a species. In order to distinguish between the scenarios, we use the notation of Ref.~\cite{Graesser:2011wi} to define the asymmetric ratio for a particle species as
\begin{equation}
r_i \equiv Y_{i}^{-}/Y_{i}^{+} {\qquad \rm~with~\quad } 0 < r_i \leq 1\,,
\end{equation}
where $+(-)$ denotes the particle (antiparticle). Here, the upper (lower) limit in $r$ signifies that the species is completely symmetric (asymmetric) and $\Omega_{\rm DM} \propto Y_i^+ + Y_i^-$. The asymptotic asymmetric ratio of a species $i$ with mass $m_i$ can be written as \cite{Graesser:2011wi}
\begin{equation} \label{eq:asymr}
r_i^{(\infty)}\simeq \exp\left[-\sqrt{\frac{\pi g_\ast}{45x_f}}\,M_{\rm Pl}\,\braket{\sigma v}_i\,\eta_D\, m_i \right],
\end{equation}
where $\braket{\sigma v}_i$ is the thermally-averaged annihilation cross section, $M_{\rm Pl}\simeq1.2\times10^{19}$ GeV, $m_p $ is the mass of the proton and $x_f \equiv m_i/T_\ast \sim20$ is the mass over freeze-out temperature ratio. In the following, we use $r_i^{(\infty)} \equiv r_i$ to alleviate notation. As shown in Ref.~\cite{BasiBeneito:2022qxd}, different regimes appear:
\begin{itemize}
\item For $r_i < 10^{-2}$, the behaviour of DM is highly asymmetric (A).
\item In the range, $10^{-2}<r_i<0.9$,  DM behaves as partially asymmetric (PA). 
\item For $r_i > 0.9$, DM is highly symmetric (S).
\end{itemize}
In the rest of the paper we adopt these ranges to define the nature of DM. Further classification can be made on the basis of the dominant production mechanism that determines the asymptotic nature of the dark matter, be it freeze-out (FO), early decays from freeze-in (FI), or late decays (LD). In this work, we aim to focus exclusively on scenarios where the asymmetry is directly involved in reproducing the relic abundance. For this goal, we propose a complete model that displays the different roles played by the asymmetries.

\section{A model for neutrino masses, the baryon asymmetry, and dark matter}{\label{sec:model}}

Our initial hypothesis for the construction of a model that explains dark matter, neutrino masses and the baryon asymmetry in a common framework is that the symmetry baryon minus lepton number, $B-L$ (under which all SM quarks have charge 1/3 and all SM leptons have charge -1), plays a key role. This is one of the best-motivated symmetries beyond the SM: it is accidental and anomaly-free in the SM, and when gauged it requires the presence of three sterile neutrinos (which in turn naturally generate active neutrino masses) and is easily embedded in GUTs. When considering a complex asymmetric dark sector, however, this symmetry is not enough, and we require extra U(1)s to forbid Majorana masses and some interaction terms, as well as to annihilate the symmetric components.

Therefore, we augment the SM gauge group by 3 new gauge $U(1)$ symmetries: $U(1)_{B-L}$ , and the dark product $U(1)_D\otimes U(1)_X$. We add three right-handed neutrinos $N_R$ (RHNs) to cancel the gauge anomalies of the first one. We also extend the particle content by the addition of two \textit{Dirac} \textit{dark} fermions, $\psi_0$ and $\chi_0$, and 3 scalars $\sigma,\,S$ and $\phi$. The gauge bosons associated with the three new groups are $Z_{B-L}^0,Z_{D}^{0},A^{'0}$, with gauge couplings $g_{B-L},g_D$ and $g_X$, respectively. The quantum numbers of the new fields are summarised in Table~\ref{tab:qnmod}. Note that all the new fields are SM singlets. The new part of the Lagrangian of the model can thus be written as
\begin{equation} \label{eq:Lagnew}
\mathcal{L}_{\rm new}=\mathcal{L}^0_{\rm \chi\psi}+\mathcal{L}^0_{\rm kin}+\mathcal{L}_{\rm int}-V(\sigma,S,\phi,H)\,,
\end{equation}
where $\mathcal{L}^0_{\rm kin}$ includes the kinetic terms of the gauge bosons and the scalars, $V(\sigma,S,\phi,H)$ is the most general scalar potential that one can write given the symmetries of the model, where $H$ is the SM Higgs doublet field, and
\begin{align}\label{eq:IntLag}
& \mathcal{L}^0_{\chi\psi}=\bar{\chi}_0(i \slashed D-m_\chi^0) \chi_0+\bar{\psi}_0(i \slashed D-m^0_\psi) \psi_0  \nonumber,\\
& \mathcal{L}_{\rm int}=-y_\nu^{\alpha i} \bar{L}^\alpha \tilde H N_R^i - y_\sigma^{ij}\sigma \overline {N^{i c}_R} {N}_R^j - y_S^{i} S\bar{N}^i_R\chi_0 - y_\phi \phi\bar{\psi}_0\chi_0+ {\rm H.c.}\,.
\end{align} 
Here $m^0_{\chi,\psi}$ are bare mass terms of the dark fermions, while the indices $\alpha=e,\mu,\tau$ and $i=1,2,3$ run over the generations of leptons and right handed neutrinos, respectively. We use $\tilde{H}=i\sigma_2H^*$. Note that $y_\nu$ is a $3\times 3$ general complex matrix, $y_\sigma$ is a $3\times 3$ complex symmetric matrix, $y_S$ is $3$-component vector and $y_\phi$ is a complex number. However, several phases are unphysical. Four phases of $y_\sigma$ may be removed by rephasing $\sigma$ and $N_R^i$. Two phases of $y_S$ may be removed by rephasing $S$ and $\chi_0$. $y_\phi$ can be taken real by rephasing $\phi$ or $\psi$.

\begin{table}[!htb]
\centering
\begin{tabular}{c c c c c}
\hline
Field & Spin & $U(1)_{B-L}$ & $U(1)_D$ & $U(1)_X$ \\
\hline
$N_R^i$ & 1/2 & -1 & 0 & 0 \\
$\sigma$ & 0& +2 & 0 & 0 \\ \hline
$\chi_0$ & 1/2 & -1 & 1 & 0   \\
$\psi_0$ & 1/2 & 0 & 0 & +1  \\
$S$ & 0 & 0 & -1 & 0 \\
$\phi$ & 0 & +1 & -1 & +1 \\
\hline
\end{tabular}
\caption{\label{tab:qnmod}Particle content of the model and their respective charge assignments under Lorentz and the $U(1)$ groups. The first two states correspond to the sterile neutrino sector, and the last four to the dark sector.}
\end{table}

In accordance with the framework discussed above, we take 
\begin{equation}
y_\phi \ll\, 1\,, \qquad g_X \ll 1\,,
\end{equation}
so that $\psi_0$ cannot thermalise with the SM bath. We further assume a vanishing initial abundance $n_\psi=0$, consistent with an inflationary epoch. On the other hand, the particles $\chi_0,\sigma,S,\phi$ and $N_R^i$ reach thermal equilibrium with the SM thermal bath through sizeable new gauge ($g_{B-L}$ and $g_D$) and scalar interactions. 

Note that the Lagrangian also contains the \textit{kinetic mixing} between the  $U(1)$ gauge factors. An unavoidable contribution to kinetic mixing among all $U(1)$s arises at one-loop level with $\phi$ running in the loop. Also, $\chi$ contributes in the case of $U(1)_{B-L}-U(1)_D$ mixing. Hence, the kinetic mixing $-(\kappa/2) Z_{B-L}^{\mu\nu}Z_{D,\mu\nu}$ is naturally of the order of $\kappa\gtrsim g_Dg_{B-L}/(16\pi^2)\sim10^{-3}\,g_{B-L}\,g_D$. The kinetic mixing of $U(1)_{X}$ with the other $U(1)$ factors is $\gtrsim g_Xg_i/(16\pi^2)$. Since $g_X\ll1$ by assumption, this contribution can be safely ignored. Finally, a kinetic mixing among $Z_{B-L}$ and the SM $U(1)_Y$ gauge boson is generated through a loop of SM quarks and leptons, of the order $\gtrsim g_Yg_{B-L}/(16\pi^2) \sum_{i=q,l}^{}Y_i(B-L)_i$. Notice that, given the conservation of $U(1)_{\rm em}$, the photon does not couple to the $B-L$ current. In any case, the bounds on the kinetic mixing are not relevant for the range of parameters that we consider in the paper.

Finally, let us remark that the necessary ingredients for the model to work could also be achieved with a global $U(1)_{B-L}$ \cite{Escudero:2016tzx,Coito:2022kif}, explicitly violated by right-handed neutrino masses, since: 
i)  $\chi$ is Dirac in nature because of the gauge $U(1)_D$, 
ii) it has sizeable interactions with the SM (with $N_R$, $y_S$) to thermalise, and iii) it can undergo efficient annihilations due to the $U(1)_D$ gauge group. However, such scenario is not as theoretically appealing as the gauged $B-L$ version that we consider, which demands the existence of 3 right-handed neutrinos. In the following subsections, we discuss the scalar, gauge, fermionic and dark sectors in detail.

\subsection{The scalar sector and spontaneous symmetry breaking}{\label{sec:scalar}}

The details of the full scalar potential $V(\sigma,S,\phi,H)$ are quite involved as one can write quadratic, quartic and mixed quartic terms for each combination of the scalar fields $\sigma,S,\phi$ and the SM Higgs doublet $H$. However, without entering into the details of the scalar potential, we can safely assume that there is a region of the parameter space in which $\sigma$ takes a large vev, $\med{\sigma}=v_{B-L}\gtrsim 10^{11}$ GeV, which breaks the $U(1)_{B-L}$ symmetry by 2 units and generates Majorana masses for the sterile neutrinos as well as a mass for the $U(1)_{B-L}$ gauge boson, $Z_{B-L}$. The value $v_{B-L}\gtrsim 10^{11}$ GeV is chosen so that the lightest RHN mass ($M_{N_1}$) safely obeys the equivalent Davidson-Ibarra lower bound \cite{Davidson:2002qv} to achieve thermal Leptogenesis in the model~\cite{Falkowski:2011xh}, but any larger value will not change the following analysis and conclusions. Therefore, an asymmetry can be generated once the inverse decays of $N_1$ go out of equilibrium \cite{Iso:2010mv,Biswas:2017tce}. Note that we take $M_{Z_{B-L}}>M_{N_1}$ and $m_\sigma>M_{N_1}$, so that the heavy $Z_{B-L}$ gauge boson and the radial component of $\sigma$ are therefore naturally very heavy and decay fast into quark and leptons.

The other scalar, $\phi$, takes a much smaller vev than that of $\sigma$, i.e., $\med{\phi}=v_\phi\ll v_{B-L}$, which breaks $U(1)_D\otimes U(1)_X\to U(1)_{X+D}$. After symmetry breaking, we can write $\phi(x)=v_\phi+{\varphi(x)}/{\sqrt{2}}$. The overall symmetry breaking pattern of the model can then be represented as
\begin{equation} \label{eq:symbr}
U(1)_{B-L}\otimes U(1)_{D}\otimes U(1)_X \overset{\med{\sigma}}{\longrightarrow}U(1)_{D}\otimes U(1)_X\overset{\med{\phi}}{\longrightarrow} U(1)_{X+D}\,.
\end{equation}
The only fields that are charged under the unbroken $U(1)_{X+D}$ symmetry are the fermions $\chi_0,\psi_0$ with charge +1 and the scalar $S$ with charge -1. We discuss in Section~\ref{sec:darkmatter} the consequences of this for the DM stability\footnote{By construction, the model does not require imposing a discrete symmetry to ensure the stability of the DM components.}.

A phenomenologically relevant parameter is the mixing between the SM Higgs boson $h$ (coming from $H=(v_{\rm EW}+h)/\sqrt{2}$) and the scalar $\varphi$, which can be characterised by the mixing angle $\theta$. This is directly related to the mixed quartic term $\lambda_{H\phi}\phi^\dagger\phi H^\dagger H$ in the scalar potential. In the rest of the paper, we assume that this mixing angle is small ($\sin\theta\simeq\theta\ll1$) so we can safely trade $h$ and $\varphi$ for the mass eigenstates.\footnote{In Appendix~\ref{sec:varphidecay}, we briefly discuss the mixing when studying the decays of $\varphi$ to SM particles. We show that even a very small mixing angle allows efficient decays.} Finally, the scalar $S$ does not obtain a vev, $\med{S}=0$,
and may therefore be a DM candidate in the model because of its charge under $U(1)_{X+D}$ (see discussion at the beginning of Section~\ref{sec:darkmatter}). 

We are interested in the regime 
\begin{equation}
M_{N_3}, M_{N_2} \gg M_{N_1}\gg m^0_\chi\gg m^0_\psi,m_S>m_\phi\,,
\end{equation}
so that the decay channels shown in Fig.~\ref{fig:cogenesis} are kinematically open. We consider values of $m_\psi^0$ and $m_S$ of similar order of magnitude, in the GeV ballpark, because we are interested in scenarios where both may contribute significantly to the DM relic abundance.

\subsection{The gauge sector}{\label{sec:gauge}}

The symmetry breaking pattern outlined in Eq.~\ref{eq:symbr} leads to one massless (two massive) gauge boson(s), which correspond to the unbroken (broken) generator(s). Up to small corrections, the masses are given by
\begin{equation}
m_{A'}^2=0\,, \quad
m_{Z_{D}}^2=2g_D^2v_\phi^2\,,\quad
m_{Z_{B-L}}^2=8g_{B-L}^2v_{B-L}^2 \,,
\end{equation}
where $A',Z_D,Z_{B-L}$ are the mass eigenstates (the full expressions can be found in Appendix \ref{sec:masslessA}). The tiny values of the parameters $g_X$, $\kappa$ and $v_\phi/v_{B-L}$ suppress the mixing between the gauge bosons so that the mass eigenstates mostly coincide with the original eigenstates. 
The massless $A'_\mu$ is decoupled from all the other fields (as $g_X\ll1$) and does not play any role in the following discussion (see Appendix~\ref{sec:masslessA} for a discussion on the bounds on a massless dark gauge boson). The mass/kinetic mixing among $Z_D$ and $Z_{B-L}$ induces an interaction of the type $Z^{\mu}_{D}J^{B-L}_\mu$, where $J^{B-L}_\mu$ is the $B-L$ current,\footnote{Note that in the absence of the mass mixing there would not be such an interaction even for $\kappa\neq0$ \cite{Heeck:2011md}.} which may lead to decays of $Z_D$ into SM quarks and leptons. 
However, the decay width is suppressed by $(v_\phi/v_{B-L})^4$ and it is subleading with respect to other decay channels:
\begin{itemize}
	\item If $m_{Z_D}>2m_\chi$, the gauge boson decays into $\bar{\chi}\chi$ or $S^\dagger S$ at tree level, or into $\varphi\varphi$ at one-loop level through a loop of $S$. This last process depends on the coupling between $\varphi$ and $S$. As we will show in Appendix~\ref{sec:varphidecay}, the particles $\varphi$ must have a fast decay to SM fermions, so this gives a lower bound on their mass, and therefore on the mass of $Z_D$.
	
	\item If $2m_{S}<m_{Z_D}<2m_\chi$, the gauge boson decays mainly as $Z_D\to S^\dagger S$. The decays have width $\sim\mathcal{O}(g_D^2m_{Z_D}/100)$ and are very fast in the relevant temperature regime ($T<v_\phi$).\footnote{We ignore the possibility $m_{Z_D}\gtrsim 2m_{S}$ in which the phase space of the decay closes.}
	\item If $2m_{S}>m_{Z_D}>2m_\varphi$, the gauge boson can only decay into $2\varphi$, see the first point. 
\end{itemize}

In the rest of the paper we focus on the second scenario, namely  $2m_{S}<m_{Z_D}<2m_\chi$.

\subsection{The fermionic sector}{\label{sec:fermion}}

We first discuss the generation of tiny neutrino masses via the type-I see-saw mechanism \cite{Minkowski:1977sc,Yanagida:1980xy,Gell-Mann:1979vob,Mohapatra:1979ia,Glashow:1979nm,Schechter:1980gr}. The breaking of $U(1)_{B-L}$ gives masses to the heavy RHNs, $M_{N_i} \sim y_\sigma^i \,v_{B-L}\gtrsim 10^{11}$ GeV. In the following we use $M_{N_1}\lesssim v_{B-L}$, i.e., we take $y_\sigma^1\lesssim1$. The masses for the active neutrinos are given by the seesaw expression,
\begin{equation}
m_\nu = -m_D\,M^{-1}_{N}\, m_D^{T}\,,
\end{equation}
with $m_D=y_\nu v_{\rm EW}/\sqrt{2}$. For the dark fermions, once $\phi$ obtains a vev, a mixing is induced between $\chi_0$ and $\psi_0$ due to the Yukawa coupling $y_\phi$, see Eq.~\eqref{eq:IntLag}. We define the fermion mixing parameter as
\begin{equation}
\epsilon_f\equiv \frac{y_\phi v_\phi}{m^0_\chi-m^0_\psi}\,,
\end{equation}
where $y_{\phi}\ll 1$ and typically $v_{\phi}< m^0_{\chi}$. Thus, the mixing parameter is highly suppressed, $\epsilon_f\ll1$ (we remind that we are assuming hierarchical masses, i.e., $m^0_\chi\gg m^0_\psi$). Upon diagonalising the fermionic sector, we have verified that the mass eigenstates $(\chi,\psi)$ mostly coincide with the original eigenstates $(\chi_0,\psi_0)$, while the corrections to the masses are $\mathcal{O}(\epsilon_f^2)$ and therefore negligible, see Appendix \ref{sec:fermionmix} for more details. Therefore, in the following we drop the subscripts 0, trading the original masses/fields for the physical ones. 

Finally, let us mention that no Majorana masses for $\chi$ or $\psi$ are generated due to the preserved gauge $U(1)_{X+D}$, due to the fact that $S$ does not take a vev. Higher dimensional operators may  be written at dimension 6 and 8 for $\chi$ and $\psi$, such as $\overline{\chi^c} \chi \sigma SS$ and $\overline{\psi^c} \psi \sigma SS \phi^\dagger \phi^\dagger$.
However, as $S$ does not take a vev, these operators do not generate Majorana masses.

\section{Dark matter components}\label{sec:darkmatter}

Recall that the only fields that are charged under the remnant $U(1)_{X+D}$ symmetry are $\chi,\psi$ (with charge $+1$) and $S$ (with charge $-1$). As $m_\chi\gg m_\psi,m_S$, only $\psi,\,S$ may be stable. The $X+D$ charge is preserved in decays of the type $\psi \rightarrow S^\dagger+ P$ or the opposite $S\rightarrow \bar{\psi}+ P$, where $P$ is some uncharged state. Therefore, in principle the lightest among $\psi$ and $S$ is stable and would be the only DM candidate.

However, the decay $\psi \rightarrow S^\dagger+ P$ (or the opposite) is suppressed by the masses of the right-handed neutrino $N_1$ and $\chi$ in the propagators. At low energies, $E\ll m_\chi \ll M_{N_1}$, one can integrate out both the particles and study the decay in terms of higher-dimensional operators. As we analyse below in Section~\ref{sec:pheno}, $P$ corresponds to active neutrinos in the model, i.e., $P=\nu$, yielding a monochromatic neutrino line from $\psi$ decays~\cite{Palomares-Ruiz:2007egs,Bell:2010fk,Garcia-Cely:2017oco,ElAisati:2017ppn,Coy:2021sse}. In that section we show that in a broad region of the parameter space, the decays are suppressed on cosmological timescales and obey current limits, and therefore both particles ($\psi$ and $S$) contribute significantly to the DM relic abundance. The scenario in which the decay is fast enough and there is only 1 DM candidate has been studied extensively in the literature, so we will not discuss it here. Hence, we focus on the two-DM scenario. The conservation at low energies of the  $U(1)_{X+D}$ symmetry yields the constraint
\begin{align} \label{eq:check}
0=Q_{X+D}&= \eta_\psi Q_\psi + \eta_S Q_S \nonumber\\
& \Longrightarrow\, \eta_\psi =  \eta_S\,,
\end{align}
where in the last step we used that the $U(1)_{X+D}$ charges are given by $Q_\psi =-Q_S=1$. First, we discuss the generation of the asymmetry and then the DM production. 

\subsection{Asymmetric dark matter via cogenesis}\label{sec:adm}

In the following, we assume a hierarchical scenario $M_{N_{2,3}}\gg M_{N_1}$, which in turn implies $y_\sigma^{2,3} \gg y_\sigma^{1}$. The simultaneous generation of lepton  and dark sector asymmetries takes place at a high scale ($T \sim M_{N_1}$) via the decays of the lightest RHN, $N_1$, into the two channels (the asymmetry generated in the decays of $N_{2,3}$ are washed out by $N_1$ interactions), as shown in Fig.~\ref{fig:cogenesis}: 
\begin{enumerate}
	\item $N_1\to LH$ generates a lepton asymmetry, $\eta_L$, which is later reprocessed into a baryon asymmetry $\eta_B$ by sphalerons.  This is the case of Type-I thermal leptogenesis, well studied in the literature (see Ref.~\cite{Davidson:2008bu} for a review), but with extra contributions, see below.
	\item $N_1\to \chi S$  generates an asymmetry in the dark sector, $\eta_D$, analogous to the lepton asymmetry. 
\end{enumerate}
Here, we focus on the regime where the asymmetry generated by $N_1$ decays into $S$ is not washed out, and is comparable to the asymmetry in $\chi$, i.e., $\eta_\chi \sim \eta_{S} \sim \eta_D$. In order to generate an asymmetry, CP needs to violated. The CP asymmetry generated in the decays of $N_1$ can be written as
\begin{align}\label{eq:cpasym}
\varepsilon_{L} = \sum_\alpha \frac{\Gamma_{N_1 \r L_\alpha H} - \Gamma_{N_1 \r \bar{L}_\alpha H^\dagger}}{\Gamma_{N_1}}\,,\quad
\varepsilon_\chi = \frac{\Gamma_{N_1 \r \chi S} - \Gamma_{N_1 \r \bar{\chi}S^\dagger}}{\Gamma_{N_1}}\,,
\end{align}
where $\Gamma_{N_1} = ((y_\nu^{} y_\nu^\dagger)_{11} + |y^1_S|^2)M_{N_1}/(8\pi)$ is the total tree-level decay width of $N_1$ and $y_S^1$ is the relevant Yukawa coupling for $N_1\to \chi +S$. In the following we refer to it as $y_S$. 

However, due to CPT invariance, no asymmetry can be generated at the tree level. CP violation arises via the interference of tree-level and one-loop level decay (vertex and self-energy corrections) amplitudes, which depends on the imaginary part of the product of the Yukawas involved. Therefore, for this asymmetry to be non-zero, we require at least two distinct phases that come from the Yukawas $y_\nu$ and $y_S$, so the couplings to the heavier neutrinos $N_{2,3}$ are important. Further, having an imaginary part in the internal loop contribution demands that the would-be decay products can be produced on-shell (i.e., the optical theorem), which is easily satisfied as we take $M_{N_1} \gg m_{\chi,S,L,H}$. 

In this cogenesis scenario, it can be seen that both $\varepsilon_L$ and $\varepsilon_\chi$ depend on $y_\nu$ and $y_S$, as the dark sector particles ($\chi, S$) are involved in the one-loop self-energy correction for $N_1 \r LH$ and vice versa. Thus, the ratio of the decay asymmetries depends on the ratio of the couplings $y_\nu$ and $y_S$ and may be correlated with the branching ratio of $N_1$ decay in each sector (${\rm Br}_L$ and ${\rm Br}_\chi$). This relates neutrino mass generation to the baryon and DM abundances.

In Ref.~\cite{Falkowski:2011xh} it has been shown that the dark asymmetry $\eta_D$ can be quite different from the visible one $\eta_B$ (contrary to the ADM models that predict $\eta_B \sim \eta_D$) due to different branching ratios, washout effects and transfer between the sectors via inverse decays and $2 \leftrightarrow 2$ scatterings. The asymptotic asymmetries for the two sectors can be written as
\begin{align}
\eta_L^{\infty} &= \varepsilon_L\, \xi_L\, Y_{N_1}^{\rm eq}(T \ll M_{N_1}) \simeq 2.6 \times 10^{-10}\,,\nonumber\\
\eta_\chi^{\infty} &= \varepsilon_\chi\, \xi_\chi\, Y_{N_1}^{\rm eq}(T \ll M_{N_1})  \,,
\end{align}
where $\xi_L\,(\xi_\chi)$ is the leptonic (dark) efficiency parameter characterising the effects of washout and transfer interactions, and $Y_{N_1}^{\rm eq}(T \ll M_{N_1})=135 \zeta(3)/4\pi^4 g_\ast$ is the initial equilibrium $N_1$ yield, with $g_\ast \simeq 106.75$, the number of relativistic degrees of freedom. The numerical value of $\eta_L^{\infty} $ is selected to match the observed baryon asymmetry, i.e., $\eta_B = (28/79)\,\eta_L$, generated via sphaleron processes. Note that the presence of $N-Z_{B-L}$ interactions in the model may modify the usual picture; however, the correct value of the asymmetry may be generated~\cite{Iso:2010mv,Hambye:2012fh}.

Typically, producing the DM relic abundance in ADM models constrains the DM mass, depending on the ratio $\eta_D/\eta_B$. However, since $\chi$ is not the DM candidate in our set-up, it is not constrained to be of the order of few GeVs (for $\eta_D \sim \eta_B$) and thus may be much larger, $\sim\mathcal{O}(\text{TeV})$. In the analysis below, we consider dark asymmetries in the range $\eta_D \sim (0.1-1)\, \eta_B$, which may be achieved if ${\rm Br}_L \sim {\rm Br}_\chi$ and the Yukawa couplings of both sectors have similar hierarchies. Thus, we work with DM masses in the GeV ballpark. We refer the reader to Fig.~6 of Ref~\cite{Falkowski:2011xh}, where it is shown the order of the asymmetries that can be obtained for a given value of $M_{N_1}$ and different branching ratios.

\begin{figure}[!htb]
	\centering
	\includegraphics[width=0.48\textwidth]{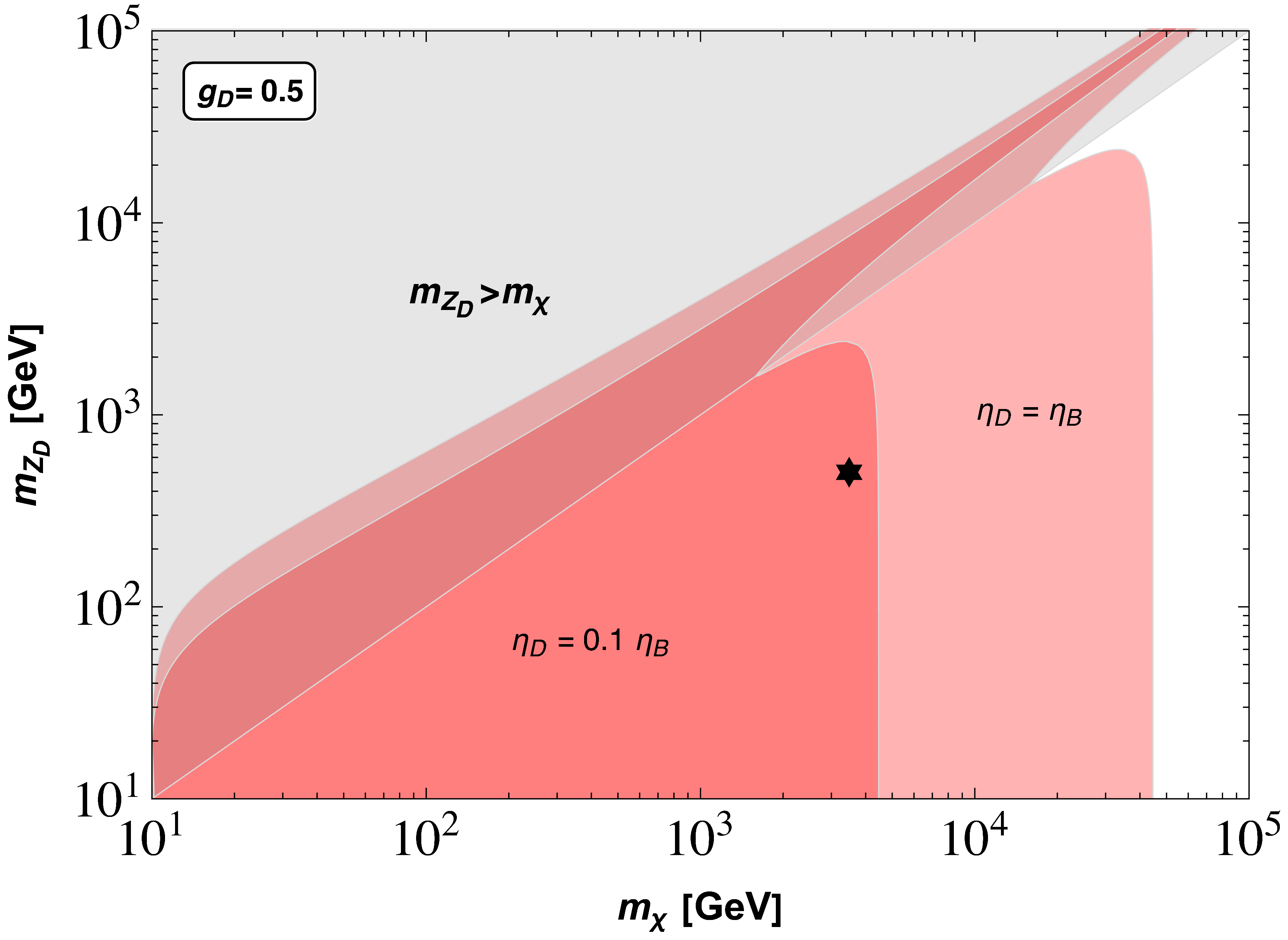}
	\includegraphics[width=0.48\textwidth]{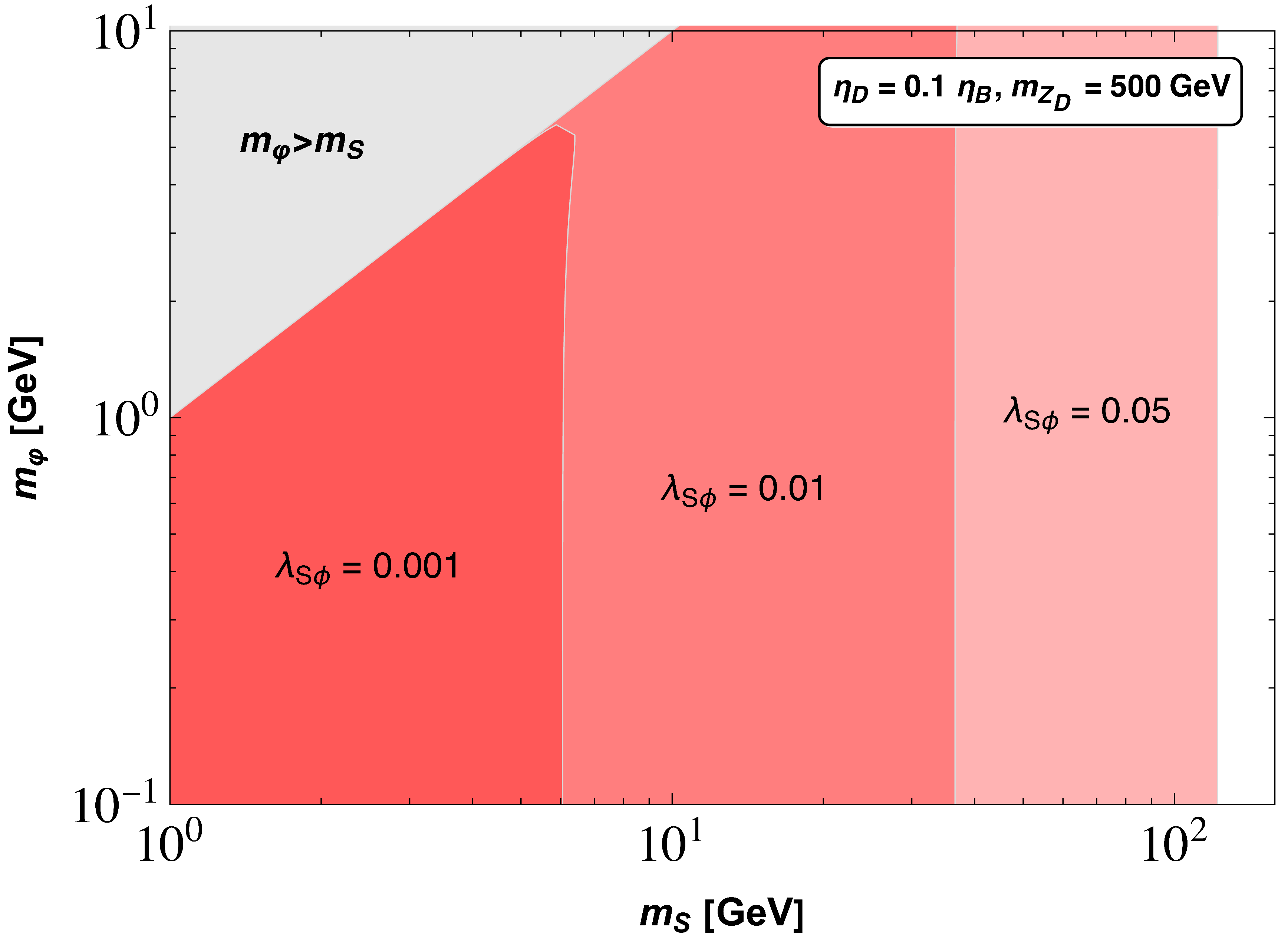}
	\caption{The red regions indicate the regions of $m_\chi, m_{Z_D}$ (\textit{left}) and $m_{S}, m_\varphi$ (\textit{right}) where the annihilations of $\chi,S$ respectively are strong enough to result in a fractional asymmetry $r_{\chi\,(S)}< 10^{-2}$. \textit{Left:} 
	The star represents the benchmark point $\{m_\chi,m_{Z_{D}}\}=\{3.5\text{ TeV}, 500\text{ GeV}\}$.  In the gray region the annihilation channel $\bar{\chi}\chi\to Z_DZ_D$ is closed. \textit{Right:} 
In the gray region, annihilations to $\varphi$ are closed. } \label{fig:chiann}	
\end{figure}

A key feature of any ADM model is the presence of an interaction that efficiently depletes the symmetric component. In our set-up, in order for the DM component $\psi$ to be asymmetric, we require that the symmetric population of $\chi$ is annihilated and only the asymmetric component participates in the late decays into $\psi$. For example, this can take place via annihilations of the form $\bar{\chi}\chi\to Z_{D}Z_{D}$. In the non-relativistic limit, $s\simeq4m_\chi^2$, the thermally-averaged cross section (considering $s$-wave annihilations) for $m_\chi > m_{Z_{D}}$ is given by
\begin{equation}
\sigma v(\bar{\chi}\chi\to Z_{D}Z_{D})=\frac{g_D^4}{16\pi m_\chi^2}\left(1-\frac{m_{Z_{D}}^2}{m_\chi^2}\right)^{3/2}\left(1-\frac{m_{Z_{D}}^2}{2m_\chi^2}\right)^{-2}.
\end{equation}
For $m_\chi > m_S$, an extra contribution comes from annihilations of the form $\bar{\chi}\chi\to Z_D\to S^\dagger S$, which opens up the region of parameter space where $m_\chi < m_{Z_{D}}$. It should be noted that there is another contribution to annihilations from the channel $\bar{\chi}\chi\to Z_{B-L}\to \bar{q}q\,(\bar{l}l)$, where $q\,(l)$ is a SM quark (lepton). However, given the large mass of $Z_{B-L}$, this turns out to be negligible. In the left plot of Fig.~\ref{fig:chiann} we show the mass ranges of $Z_D$ and $\chi$ where the latter is asymmetric, i.e., $r_\chi<10^{-2}$, see Eq.~\ref{eq:asymr}. One observes how, for larger values of the asymmetry, a larger region of the parameter space has $r_{\chi}<10^{-2}$. Note also the presence of the resonance for $m_{Z_D}\simeq 2 m_\chi$.

\subsection{Contribution of \texorpdfstring{$\psi$}{psi}}

\subsubsection{Production of \texorpdfstring{$\psi$}{psi} from \texorpdfstring{$\chi$}{chi} decays}

First, we study the dynamics of $\psi$, neglecting the contribution to the relic abundance from $S$. In this limit, $\chi$ can only decay to $\psi\phi\,(\varphi)$. We assume that the $\bar{\chi}\chi$ annihilation processes studied in the previous section are efficient enough so that $r_\chi<10^{-2}$ and the freeze-out abundance of $\chi$ is determined by its asymmetry. The production of $\psi$ is driven by decays of $\chi$: $\chi\to\psi \phi$ for $T>v_\phi$ and $\chi\to\psi \varphi$ for $T<v_\phi$. As the $h-\varphi$ mixing is small ($\theta \ll 1$), it is safe to trade $\varphi,h$ for the mass eigenstates. The production via decays can be divided into two types:
\begin{enumerate}
	\item While $\chi$ is in thermal equilibrium with the SM bath ($T>T_\ast$, $T_\ast\simeq m_\chi/20$ being the freeze-out temperature of $\chi$,), the decays are symmetric, i.e., $\psi$ and $\bar{\psi}$ are produced in equal amounts from the decays of $\chi$ and $\bar{\chi}$, that are symmetric during this period, i.e., $Y_\chi^+ \simeq Y_{\chi}^-=Y_{\chi}^{\rm eq}\gg\eta_D$. This is the usual freeze-in contribution. We also refer to these processes as \textit{early} decays.  We denote the abundance of $\psi$ particles produced by freeze-in by $Y_{\rm FI}$. This symmetric production peaks around $T\sim m_\chi>v_\phi$, so that the channel is  $\chi\to\psi \phi$. 
	\item Once $\chi$ freezes out ($T<T_\ast)$, the population of $\chi$ becomes asymmetric, i.e., $Y_\chi^+ \simeq \eta_D\gg Y_{\chi}^- \simeq r_\chi Y_\chi^+$. Such an asymmetry is then subsequently transferred to $\psi$ via decays. This is the asymmetric freeze-in contribution of $\psi$ from \textit{late} decays (LD). Since the decays occur late, i.e $T_D\ll v_\phi$, in this case the channel is $\chi\to \psi\varphi$. The populations are given by
\begin{equation}\label{eq:psiaa}
Y_{\rm LD}^+ =\frac{\eta_D}{1-r_\chi}\simeq \eta_D\,,\\ \qquad Y_{\rm LD}^-= r_\chi Y_{\rm LD}^+ \simeq \eta_D\,r_\chi\,,
\end{equation}
where in the last step we used $r_\chi<10^{-2}$.
\end{enumerate}
Therefore, the asymptotic abundance of $\psi$ and $\bar{\psi}$ can be written as
\begin{align}\label{eq:psiab}
Y_\psi^+ =\frac{Y_{\rm FI}}{2}+Y^+_{\rm LD}\simeq  \frac{Y_{\rm FI}}{2}+{\eta_D}\,,\quad Y_{\psi}^- = \frac{Y_{\rm FI}}{2}+Y^-_{\rm LD}\simeq \frac{Y_{\rm FI}}{2}+\eta_D\,r_\chi\,,
\end{align}
where in the last step we used Eq.~\eqref{eq:psiaa}. We show in the next section that Eq.~\eqref{eq:psiab} gets modified by a multiplicative factor when the contribution of the scalar $S$ is taken into account.
The three interesting cases for the asymmetric ratio are 
\begin{equation}
r_\psi\simeq 
\begin{cases}
r_\chi \hspace{1.55cm} \text{ if \,\,} \eta_Dr_\chi\gg Y_{\rm FI}\,, \\
Y_{\rm FI}/(2\eta_D) \hspace{0.23cm} \text{ if \,\,} \eta_D r_{\chi}\ll Y_{\rm FI}\ll \eta_D\,, \\
1 \hspace{1.75cm} \text{ if \,\,} Y_{\rm FI}\gg \eta_D\,.
\end{cases}
\end{equation}
We remind that, according to our definition in Section~\ref{sec:fram}, DM is asymmetric if $r_\psi<10^{-2}$, partially asymmetric if $10^{-2}<r_\psi<0.9$ and symmetric if $r_\psi>0.9$.
Hence, in order for $\psi$ to be asymmetric, not only we need to require that $r_\chi\ll1$ but also the symmetric freeze-in contribution to the abundance, $Y_{\rm FI}/2$, should be suppressed, i.e.,  $Y_{\rm FI}\ll \eta_D$. In the opposite regime, where the freeze-in production from early decays is dominant, the final $\psi$ abundance is always symmetric. The decay width is given by 
\begin{equation}\label{eq:decayw}
\Gamma_{\chi \r \psi\varphi}\simeq \frac{y_\phi^2m_\chi}{32\pi}\,\Delta^2(f_\psi,f_{\varphi})\,,
\end{equation}
where $f_\psi \equiv m_\psi/m_\chi$ and $f_\varphi\equiv m_\varphi/m_\chi$, and $\Delta(f_\psi,f_\varphi)$ is the phase space suppression factor,
\begin{equation}
	\Delta^2(f_\psi,f_\varphi)=\left(1-f_\psi^2-f_\varphi^2+2f_\psi\right)^2\left[\left(1-f_\psi^2-f_\varphi^2\right)^2-4f_\psi^2 f_\varphi^2\right]\,,
	\end{equation}
with $0\leq\Delta(f_\psi,f_\varphi)\leq1$. We have neglected the small corrections due to fermion mixing $\epsilon_f$. There is an analogous expression for $\varphi\to\phi$. Here onwards, we omit the arguments of the function $\Delta$ and we consider values of the parameters such that $m_\chi\gg m_\psi, m_\varphi$. Within this approximation $\Delta \simeq1$. Furthermore, in this limit the width is practically independent of $m_\varphi$ or $m_\phi$, so that  in the following we do not differentiate among decays into $\psi\varphi$ and $\psi\phi$. Notice that the Yukawa $y_\phi$ also leads to annihilation processes (such as $\bar{\chi}\chi\to\bar{\chi}\psi$) which would contribute to $\psi$ production. However, as long as $\Delta\simeq1$, these are subleading with respect to decays.

\begin{figure}[!htb]
	\centering
	\includegraphics[width=0.6\textwidth]{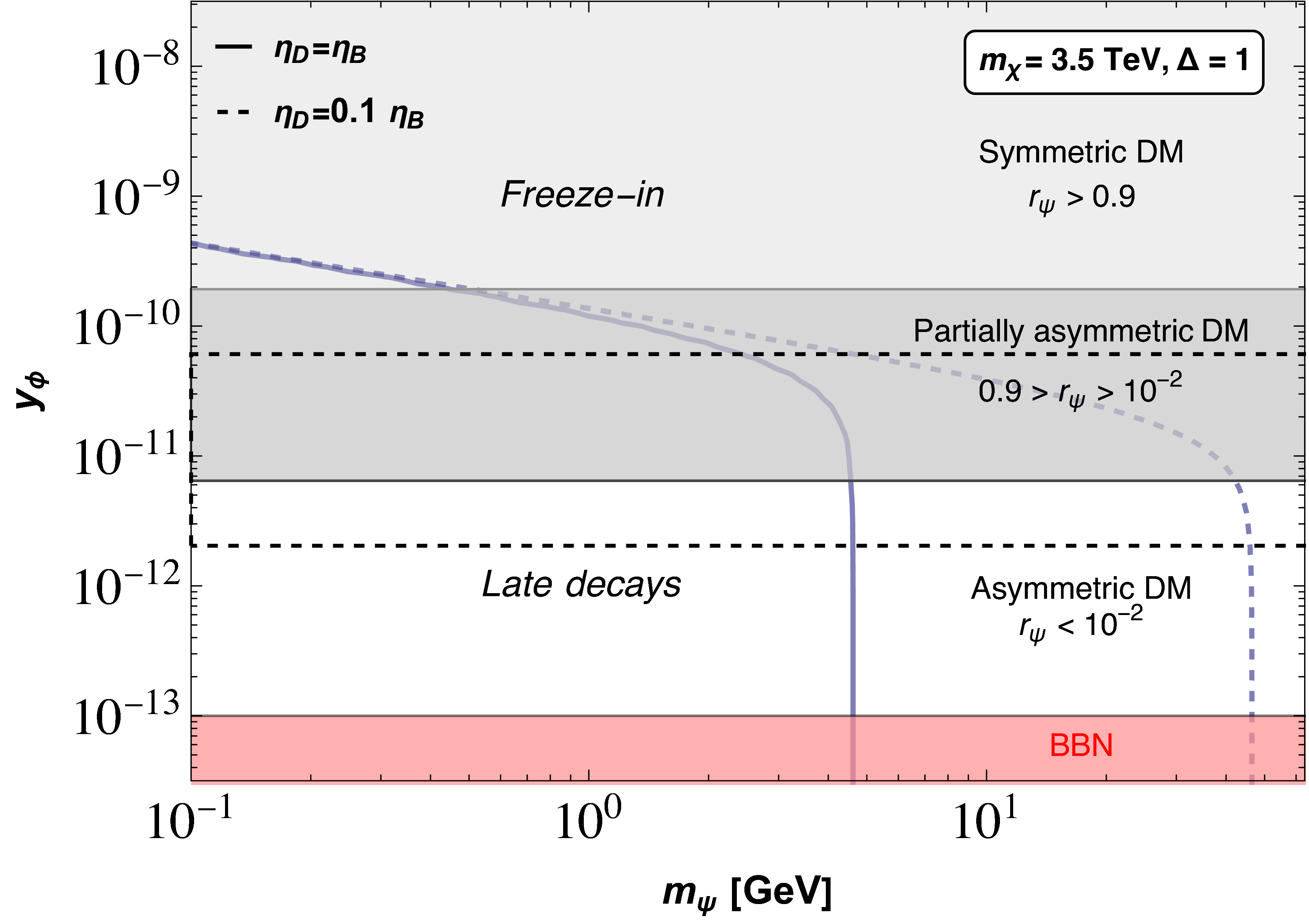}
	\caption{Parameter space for the scenario in which the DM is composed solely of $\psi$ and we assume that $S$ is light enough in each point of the plot so that its contribution to the DM abundance can be safely neglected. We distinguish three different regions depending on the nature of DM (symmetric, asymmetric or partially asymmetric), labelled by the value of $r_\psi$. In the darker gray region DM is produced through freeze-in and is symmetric. In the white region it is produced via late decays and is asymmetric. The DM relic abundance is reproduced along the blue solid (dashed) line which corresponds to $\eta_D/\eta_B=1\,(0.1)$, whereas the horizontal dashed lines indicate the shift in the regions for $\eta_D =0.1 \eta_B$.} \label{fig:dmpsi}
\end{figure}

The freeze-in contribution from early decays can be computed numerically by solving the  Boltzmann equations for $\chi$ and $\psi$. However, a very good analytic estimate is given by
\begin{equation} \label{eq:FI}
Y_{\rm FI}\simeq \frac{135}{8\pi^4}\,\sqrt{\frac{45}{\pi g_{\ast}^3}}\,\frac{\Gamma_{\chi \r \psi\varphi}\,M_{\rm Pl}}{m_\chi^2}\simeq 6\times10^{-6}\,{y_\phi^2~\Delta^2}\,\frac{M_{\rm Pl}}{m_\chi}\,,
\end{equation}
where we used that the production peaks around the mass of the heaviest particle involved in the decay, i.e., at $T\simeq m_\chi$.
The late decays of $\chi$ would instead peak at temperature $T_D$ at which $\Gamma_{\chi \r \psi \varphi}/H|_{T=T_D}\simeq 1$,
\begin{equation}\label{eq:TD}
\begin{split}
T_D&\approx y_\phi\,\Delta\,\sqrt{\frac{m_\chi M_{\rm Pl}}{32\pi}} \approx 10{\rm~ MeV}\,\Delta\,\left(\frac{y_\phi}{10^{-12}}\right)\,\left(\frac{m_\chi}{3.5\text{ TeV}}\right)^{1/2}.
\end{split}
\end{equation}
In order to get asymmetric DM, we impose the condition $Y_{\rm FI}\lesssim10^{-2}\eta_D$ (here we are assuming values of $\{g_D,m_\chi,m_{Z_{D}}\}$ such that $r_\chi\lesssim10^{-2}$) which implies
\begin{equation}\label{eq:condFI}
y_\phi\,\Delta\lesssim 6\times10^{-12}\,\left(\frac{\eta_D}{\eta_B}\right)^{1/2}\left(\frac{m_\chi}{\text{ 3.5 TeV}}\right)^{1/2}.
\end{equation}
In Fig.~\ref{fig:dmpsi}, we show the parameter space in the plane $y_\phi$ versus $m_\psi$ where the contribution of $\psi$ is dominant. We fix $m_\chi=3.5$ TeV. We highlight the regions where it is symmetric, partially-asymmetric and asymmetric, subject to the constraints discussed below.

\subsubsection{Constraints} \label{sec:constpsi}

The gauge coupling $g_X$ is constrained by long-range force experiments, $g_X\lesssim 10^{-8}$,  see Appendix~\ref{sec:masslessA}. In the rest of the paper, we take $g_X$ sufficiently small so that gauge interactions can not drive thermalisation of $\psi$ and therefore $g_X$ never plays a role. On the other side, $\psi$ could thermalise through processes involving the Yukawa coupling $y_\phi$. All these processes involve at least one $\chi$ particle (or the conjugate). The most important ones are the decays (and inverse decays) $\chi\to\psi\varphi$. The condition for non-thermalisation is $\Gamma_{\chi\r \psi\varphi }/H|_{ T\simeq m_\chi}\ll 1$. Indeed, for $T>m_\chi$, the interaction rate/Hubble ratio grows when the temperature decreases, while for $T<m_\chi$ decay and inverse decays become inefficient. Therefore, we evaluate the ratio at its maximal value, i.e., $T\simeq m_\chi$, which gives 
\begin{equation}\label{eq:ybound}
y_\phi\,\Delta\ll30\sqrt{\frac{m_\chi}{M_{\rm Pl}}}= 5\times 10^{-7}\sqrt{\frac{m_\chi}{3.5\text{ TeV}}}\,.
\end{equation}
Notice that taking the limit $\Delta\ll1$ does not help in pushing towards higher values of the Yukawa. Indeed, if $\Delta$ is small, the decay channel closes and annihilations become more relevant. Annihilations processes are independent from $\Delta$ and give a condition analogous to Eq.~\eqref{eq:ybound} with $\Delta=1$. However, as we are considering $\Delta\simeq1$, decays are more important than annihilations. 

In the case in which $\psi$ is produced mainly by late decays, we require that these are peaked much after the freeze-out of $\chi$, i.e., $T_D\ll T_*$, which ensures that the population of $\psi$ is asymmetric. This gives a stronger condition,
\begin{equation}\label{eq:ybound2}
y_\phi\,\Delta\ll0.5\sqrt{\frac{m_\chi}{M_{\rm Pl}}}= 8.5\times 10^{-9}\sqrt{\frac{m_\chi}{3.5\text{ TeV}}}\,.
\end{equation}
If $\psi$ is produced by early decays, peaked around $T_{\rm prod}\simeq m_\chi$, it behaves as cold DM as long as $m_\psi\gtrsim$ keV. However, if the production from late decays is significant, we must take into account an additional constraint. Indeed, as
we have a heavy particle ($\chi$) decaying late into a lighter stable one ($\psi$), we must check that the DM free streaming length is smaller than 0.1 Mpc, which gives \cite{Heeck:2017xbu}
\begin{align}\label{eq:WDM1}
m_{\psi}&>3.5{\rm~keV}\,\med{p/T}_{\rm prod}\,\left(\frac{10}{g_{\ast}(T_{D})}\right)^{1/3}\gtrsim  1.7{\rm~keV}\,\frac{m_\chi\,\Delta}{T_D}\,\left(\frac{10}{g_{\ast}(T_{D})}\right)^{1/3}\,,
\end{align}
where in the second step we used that the typical DM momentum at production is $\med{p}\sim m_\chi\Delta/2$, while $\med{T_{\rm prod}}=T_D$. Using Eqs.~\eqref{eq:TD} and~\eqref{eq:WDM1}, this translates into a lower bound on the coupling,
\begin{equation}
y_\phi\gtrsim 5.6\times 10^{-14}\,\left(\frac{m_\chi}{3.5{\rm~TeV}}\right)^{1/2}\,\left(\frac{5\text{ GeV}}{m_\psi}\right)\,\left(\frac{10}{g_{\ast}(T_D)}\right)^{1/3}\,.
\end{equation}
This bound only applies if \emph{i)} $\psi$ reproduces the DM relic abundance, and \emph{ii)} the production from late decays is dominant, i.e., $Y_{\rm LD}^+\gg Y_{\rm FI}$, corresponding to the vertical blue lines of Fig.~\ref{fig:dmpsi}.
We can use BBN bounds to constrain the lifetime of $\chi$. If $\chi$ did not decay, its abundance today, $\Omega_{\chi}h^2$, would be $m_\chi/m_\psi$ times the would-be DM ($\psi$) abundance. In particular, for $m_\chi\sim \mathcal{O}{\rm (TeV)}$ and would-be abundance $\eta_D\sim10^{-11}$, this leads to $\tau_\chi\lesssim(0.1-1)$ s \cite{Jedamzik:2006xz} for
\begin{equation}\label{eq:BBN}
y_\phi\Delta\gtrsim {10^{-13}}\,\left(\frac{3.5{\rm~TeV}}{m_\chi}\right)^{1/2}.
\end{equation}
Here we assumed that the decay of $\varphi$ into SM radiation (mainly hadrons) occurs instantaneously right after $\chi$ decays; otherwise, the bound should read $\tau_\chi+\tau_{\varphi}\lesssim(0.1-1)$ s. This is discussed in Appendix~\ref{sec:varphidecay}. The effects of fermion mixing are discussed in Appendix~\ref{sec:fermionmix}.

\subsection{Contribution of $S$}

\subsubsection{Production of $S$ from freeze-out}

The scalar $S$ is produced by $N_1$ decays at high temperature and shares the same asymmetry of $\chi$, i.e,  $\eta_D$. Once produced, it thermalises with the SM bath through scalar and gauge interactions and undergoes annihilations. 
In the following, we focus mainly on masses of $S$ in the $1-50$ GeV range, so that its contribution to the relic abundance may be of the same order as that of $\psi$. In this range of masses, annihilations of $S$ into $Z_D$ may be kinematically forbidden. However, if $m_\varphi<m_{S}$ (possible by tuning the coefficients of the scalar potential, e.g., $m_\varphi/v_\phi\lesssim 10^{-3}$ and $m_\varphi \sim \mathcal{O}({\rm GeV})$ or lighter), then the annihilations $S^\dagger S\to\varphi\varphi$ are allowed. Next we consider the minimal option of just the scalar portal $\lambda_{S\phi}|S|^2|\phi|^2$. The non-relativistic cross section for $S^\dagger S\to \varphi\varphi$ induced by the operator is given by \footnote{We neglect the diagram containing the self-interaction $\varphi^3$, which enters if the $\phi$ quartic coupling is not extremely small.} 
\begin{equation}
\sigma v (S^\dagger S \r \varphi\varphi)\simeq \frac{\lambda_{S\phi}^2}{32\pi m_{S}^2}\,\left(1-\frac{m_\varphi^2}{m_{S}^2}\right)^{1/2}\,\left(\frac{1-{m_\varphi^2}/{2m_{S}^2}-2\lambda_{S\phi}{v_\phi^2}/{m_{S}^2}}{1-{m_\varphi^2}/{2m_{S}^2}}\right)^2\,.
\end{equation} 
Even for moderately small values of the coupling, the cross section is significant and annihilations are strong enough to destroy the symmetric population of $S$. Hence, the symmetric population of $S$ annihilates around $T_*^{(S)}\sim m_{S}/20$, leaving only the asymmetric population, with the abundance fixed by the dark asymmetry $Y_{S}^+ \sim\eta_D$, $Y_{S}^{-}\sim r_SY_{S}\ll Y_{S}^+$, completely analogous to the computation of $\chi$ annihilations. In Fig.~\ref{fig:chiann} (right panel) we show the region of $m_\varphi$ and $m_{S}$ where $S$ is  asymmetric for different values of the coupling $\lambda_{S\phi}$ fixing $\eta_D =0.1 \eta_B$ and $m_{Z_D} \sim 500{\rm~GeV}$.

In principle, other annihilation channels for $S$ are possible, depending on the scalar potential parameters, such as $S^\dagger S\to h\to \bar{f}f$ (which is more suppressed for $m_h>m_{S}$), allowing for a larger set of possibilities. For simplicity, we focus only on annihilations into $\varphi\varphi$, which involve only one coupling $\lambda_{S\phi}$ and need not be very large, $\mathcal{O}(10^{-2})$.

\subsubsection{Production of $S$ from \texorpdfstring{$\chi$}{chi} decays}

Integrating out the lightest of the heavy sterile neutrinos, $N_1$, induces the effective interactions
\begin{equation} \label{eq:OD}
y_\nu y_S \frac{\bar{L}\tilde H S \chi}{M_{N_1}}\,,\,\,\,\qquad y_S^2 \frac{\chi^2S^2}{M_{N_1}}\,.
\end{equation}
 The former leads to $\chi\to S^\dagger \nu_L$ and $\chi\to S^\dagger \nu_L h$ decays, which compete with $\chi \r \psi\varphi$, eventually populating the $S^\dagger$ sector, whereas, the latter induces $\chi\chi\to S^\dagger S^\dagger$ annihilations. Focusing on decays, we can use the estimate (see also Ref.~\cite{Coy:2021sse})
 \begin{equation}\label{eq:decay2}
 \Gamma(\chi\to S^\dagger\nu)\approx\frac{|y_S|^2m_\chi}{32\pi}\left(\frac{m_\nu}{M_{N_1}}\right)\left(1-\frac{m_{S}^2}{m_\chi^2}\right)^2.
 \end{equation}
These decays are peaked around $T_D^{(S)}$ defined as $\Gamma_{\chi\to S^\dagger\nu}/H|_{T=T_{D}^{(S)}}=1$ (analogous to the decays into $\psi$ computed earlier) and given by
\begin{equation}
T_D^{(S)}\simeq{|y_S|}\sqrt{\frac{m_\nu m_\chi M_{\rm Pl}}{32\pi M_{N_1}}}\simeq \text{ MeV} \left(\frac{|y_S|}{10^{-3}}\right)\left(\frac{m_\nu}{0.05 \text{eV}}\frac{m_\chi}{3.5\text{TeV}}\frac{10^{11}\text{GeV}}{M_{N_1}}\right)^{1/2}\,.
\end{equation}
Since for the values of the parameters $T_D^{(S)}<T_*$, the only important decay is $\chi\to S^\dagger\nu_L$, while the conjugate process is irrelevant as the population of $\bar{\chi}$ after freeze-out is negligible (recall that we are in the region of the parameter space in which $r_\chi<10^{-2}$).\footnote{Also in this case we can distinguish between the late decays, peaked at $T_D^{(S)}$, and the early decays at $T>T_*$. The latter produce a symmetric population of $S$ and $S^\dagger$ from the decays of $\chi$ and $\bar{\chi}$ (the analogue of the freeze-in population of $\psi$, peaked at $T\simeq m_\chi$). However, this symmetric population thermalises and undergoes annihilations leaving no imprint in the final abundance.}

We can parametrise the dominant decay channel of $\chi$ by defining the ratio of branching ratios
\begin{equation} \label{eq:R}
R\equiv \frac{{\rm Br}(\chi\to S^\dagger\nu)}{{\rm Br}(\chi\to\psi \varphi)}\sim \frac{|y_S|^2}{y_\phi^2} \frac{m_\nu}{M_{N_1}}\,,
\end{equation}
where in the last step we used Eqs.~\eqref{eq:decayw} and~\eqref{eq:decay2}. The decay of $\chi$ into $\psi$ is the dominant channel, i.e., $R\ll1$, for
\begin{equation}\label{eq:y1b}
|y_S|\ll\left(\frac{y_\phi}{2\times 10^{-11}}\right)\,\left(\frac{M_{N_1}}{10^{11}\text{GeV}}\right)^{1/2}\,\left(\frac{0.05\text{eV}}{m_\nu}\right)^{1/2}\,.
\end{equation}
Notice that the contribution of the $3$-body decays $\chi\to\nu h S^\dagger$ should be similar to the $2$-body ones, as the decay rate gets suppressed by the phase space factor while at the same time it has an enhancement $(m_\chi/v)^2\sim 200$ (eventually it can become dangerous for $m_\chi\gg$ TeV). For a generic value of $R$, the probability to decay into $\psi$ is $1/(1+R)$ while into $S^\dagger$ is $R/(1+R)$. Therefore the abundances of $\psi,\bar{\psi}$  in Eq.~\eqref{eq:psiab} get divided by $(1+R)$.

Concerning $S$, we can distinguish two possibilities: 
\begin{enumerate}
\item If $T_D^{(S)}<T_*^{(S)}<T_*$, at the time where decays into $S^\dagger$ peak, the latter has already decoupled from the thermal bath with an asymmetric abundance $Y_{S}=\eta_D$ and therefore the $S$ and $S^\dagger$ population can not annihilate. Then the abundances of $S,S^\dagger$ are
\begin{equation} \label{eqYS}
Y_{S} \equiv Y^+_{S}=\eta_D,\, \qquad Y_{S^\dagger}\equiv Y^-_{S}=\frac{R}{1+R}\,\eta_D\,,
\end{equation}
where we assume that  $r_S<10^{-2}$ and we have ignored it for simplicity. This corresponds to
\begin{equation}\label{eq:LDS11}
|y_S|\lesssim 0.1\left(\frac{m_{S}}{\text{GeV}}\right)\,\left(\frac{M_{N_1}}{10^{11}\text{GeV}}\right)^{1/2}\,\left(\frac{0.05\text{eV}}{m_\nu}\right)^{1/2}\,\left(\frac{3.5\text{TeV}}{m_\chi}\right)^{1/2}.
\end{equation}

\item If $T_*^{(S)}<T_D^{(S)}<T_*$, the decays produce a population of $S^\dagger$, while $S^\dagger S$ annihilations are still efficient. As a result there is a partial washout of $\eta_S$, which gets reduced to $\eta_D/(1+R)$. Therefore, at $T<T_*^{(S)}$ a population of $S$ decouples with abundance
\begin{equation}\label{eqYS2}
Y_S^+ = \frac{1}{1+R}\,\eta_D\,, \qquad Y_S^-\ll Y_S^+\,,
\end{equation} 
where we assume $r_S<10^{-2}$.
Notice that for $R\ll1$ Eqs.~\eqref{eqYS} and \eqref{eqYS2} are equivalent as the decays are irrelevant.
\end{enumerate}

\subsubsection{Constraints}

We can constrain the value of $|y_S|$ for the case in which $R>1$. As for this choice $\chi$ decays mostly into $S^\dagger$, we must impose that the decay occurs before BBN, in analogy with Section~\ref{sec:constpsi}. Using the decay rate in Eq.~\eqref{eq:decay2} we find that the BBN bound translates into the constraint
\begin{equation}
|y_S|\gtrsim10^{-3}\left(\frac{0.05\,\text{eV}}{m_\nu}\frac{M_{N_1}}{10^{11}\,\text{GeV}}\frac{3.5\,\text{TeV}}{m_\chi}\right)^{1/2}.
\end{equation}
For $R<1$ we can choose smaller $y_S$ while BBN constrains the value of $y_\phi$, see Eq.~\eqref{eq:BBN}. 

We also find that for $R>1$ the model is characterised by an interesting feature: the decays $\chi \to S^\dagger\nu_L$, which occur before BBN and neutrino decoupling (around $T\sim \mathcal{O}(10 {\rm~MeV})$ for the choice of parameters we adopted and $|y_S|\sim 10^{-2}$), generate also an asymmetric population of neutrinos,
 \begin{equation}
 \Delta \eta_\nu=\frac{R}{(1+R)}\,\eta_D\,,
 \end{equation}
 which is maximal ($\approx\eta_D$) for large $R$. Therefore, the neutrino population is more asymmetric than in the standard case, as this new contribution sums up to the usual leptonic asymmetry generated earlier on by leptogenesis. Note, however, that since these decay processes occur below the scale of electroweak symmetry breaking, this leptonic asymmetry is not transferred to the baryonic one. 
 
Finally, if $R>1$ (and $T_D^{(S)}<T_*^{(S)}$), the $S^\dagger$ population that arises from late decays of $\chi$ may be warm. In such a case, if a significant fraction of the DM was made by this $S^\dagger$ population, constraints from free streaming length (see Eq.~\ref{eq:WDM1} with $m_\psi\to m_S$ and $T_D\to T_D^{(S)}$) would give $|y_S|\gtrsim 5 \text{ MeV}/m_S$. However, even in the cases in which the contribution of $\psi$ DM is negligible, the DM is composed by a mixture of $S$ (produced by freeze-out, always cold) and $S^\dagger$, where $Y_S^+\geq Y_S^-$ (the equality applies if $R\gg1$). This corresponds to a mixture of cold/warm DM, which in general is less constrained than the warm DM case \cite{Boyarsky:2008xj}. 
 
In the following, we discuss the scenarios arising due to the different dynamics of the dark components in the model. It can be checked that Eq.~\ref{eq:check} is always fulfilled.

\section{Dark matter relic abundance}\label{sec:dmrelic}

In this section, we discuss the possible scenarios that can be distinguished depending on the values of the parameters $y_\phi$ and $R$ (which can be rephrased as a function of $y_\phi$ and $|y_S|$), while satisfying all the constraints. We restrict our analysis to the region of the parameter space in which:
\begin{enumerate}
	\item Both $\psi$ and $S$ are stable, see the discussion in Section~\ref{sec:pheno}.
	
	\item $\bar{\chi}\chi\to Z_DZ_D$ annihilations are efficient enough so that $r_\chi<10^{-2}$, i.e., the late decays of $\chi$ are always asymmetric.
	
	\item $S^\dagger S\to \varphi\varphi$ annihilations are efficient enough so that $r_{S}<10^{-2}$, i.e., only the asymmetric population of $S$ survives the annihilations.
\end{enumerate}
Regarding the Yukawa parameter $y_\phi$, we consider the range $10^{-13}\lesssim y_\phi\lesssim 5\times 10^{-7}$, where the lower bound comes from BBN (only applicable if $R<1$), while the upper one comes from the requirement that $\psi$ does not thermalise. Regarding the DM component $\psi$, we identify its nature depending on the value of $y_\phi$. In the two limiting cases, we have
\begin{itemize}
	\item \underline{\textit{Asymmetric $\psi$:}} If $y_\phi\lesssim6\times10^{-12}\sqrt{\eta_D/\eta_B}$, the freeze-in population of $\psi$ from early decays is negligible, while the late decays are dominant, resulting in an asymmetric population $Y_\psi=\eta_D\gg Y_{\rm FI}$.
    \item \underline{\textit{Symmetric $\psi$:}} If $y_\phi\gtrsim2\times10^{-10}\sqrt{\eta_D/\eta_B}$, the $\psi$ population is dominated by freeze-in (early decays), $Y_{\rm FI}\gg \eta_D$. The resulting $\psi$ population is symmetric.
\end{itemize}
At the same time, we identify the nature of the DM component $S$ according to the value of the parameter $R$. The two limiting cases are:
\begin{itemize}
	\item \underline{\textit{Asymmetric $S$:}} If $R\ll1$ (corresponding to $|y_S|\ll y_\phi/(2\times10^{11})$), $\chi$ decays into $S^\dagger$ are negligible and the abundance of $S$ is determined by the asymmetry via freeze-out, i.e., $Y_S=\eta_D$.
	\item \underline{\textit{Symmetric $S$:}} If $R\gtrsim\mathcal{O}(10)$, $\chi$ mostly decays into $S^\dagger$, so that the final abundance of $S,S^\dagger$ population is the same, $Y_S=Y_{S^\dagger}=\eta_D$  and DM is symmetric. The interesting feature of this scenario is that the DM abundance is set by the asymmetry $\eta_D$ even though the nature of DM is symmetric. Furthermore, given the large value of $R$, the production of $\psi$ is negligible. Therefore the scenario leads to practically only 1 DM component with abundance completely fixed by the asymmetry: $Y_S+Y_{S^\dagger} =2\eta_D$, leading to the prediction
	\begin{equation} \label{eq:mass6}
	m_{S}\simeq2.5\text{ GeV}\, \left(\frac{\eta_B}{\eta_D} \right)\,.
	\end{equation}
	This scenario (scenario 6 according to our classification in Table~\ref{tab:sum}) is phenomenologically interesting as it leads to an enhanced indirect detection signal compared to the usual freeze-out case.  
\end{itemize}
\begin{table}[!htb]
	\centering
	\small
	\begin{tabular}{c c c c c c}
		\hline
		\textbf{Sc.} & $\psi$ population & $S$ population &
		$10^{-10}y_\phi/\sqrt{\eta_D/\eta_B}$  & $R$ & $T_D^{(S)}/T_*^{(S)}$ \\
		\hline
		\textbf{1} & Asymmetric & Asymmetric &
		$\leq 0.06$ & $\ll 1$& Any \\
		\color{red}{\textbf{2}}
		 & Asymmetric & Partially Asymmetric & $\leq0.06$ & $\mathcal{O}(1)$ & $<1$ \\
		\textbf{1-2}
		 & Asymmetric & Asymmetric &
		$\leq0.06$ & $\mathcal{O}(1)$ & $>1$  \\		
		\color{blue}{\textbf{3}}
	& Partially Asymmetric & Asymmetric	& $0.06-2$ & $\ll 1$& Any \\
		\color{green}{\textbf{4}}
		& Partially Asymmetric & Partially Asymmetric & $0.06-2$ & $\mathcal{O}(1)$ & $<1$ \\
		\textbf{3-4}
	& Partially Asymmetric & Asymmetric	& $0.06-2$ & $\mathcal{O}(1)$  & $>1$ \\
		\color{purple}{\textbf{5}} & Symmetric & Asymmetric &
		$\gtrsim 2$ & $\ll 1$  & Any \\
		\hline
		\color{orange}{\textbf{6}} & Negligible & Symmetric &
		$y_\phi\lesssim5\times10^{-7}$ & $\gtrsim \mathcal{O}(10)$  & $<1$ \\	
		\hline
	\end{tabular}
	\caption{\label{tab:sum} Classification of the different scenarios depending on the nature of $\psi/S$ population and the values of $y_\phi$ (in units of $\sqrt{\eta_D/\eta_B}$) and $R$. The values of the parameters are fixed: $m_\chi=3.5$ TeV, $m_{Z_D}=500$ GeV, $g_D=0.5$, $M_{N_1}=10^{11}$ GeV, and $m_\nu=0.05$ eV. We take $y_\phi<5\times10^{-7}$ to avoid thermalisation of $\psi$. In Scenarios 2, 4 and 6 the Yukawa satisfies $|y_S|\lesssim0.1(m_S/\text{GeV})$. In the Mixed Scenarios 1-2 and 3-4 we have $|y_S|>0.1(m_S/\text{GeV})$. Scenarios 1, 3 and 5 give the same result independently of this condition.}
\end{table}
The intermediate regimes give rise to new possibilities, such as partially-asymmetric DM for the different components. The classification of scenarios depending on the nature of the $\psi/S$ population and the values of $y_\phi$ and $R$ is shown in Table~\ref{tab:sum}. The scenarios are discussed in detail in Appendix~\ref{app:scenarios}, where we also provide the classification on the basis of dominant production mechanism and asymptotic nature of both dark matter components in Table~\ref{tab:scesum}. For all the scenarios, one can use the following expressions for the individual and total relic abundance,
\begin{align}\label{eq:DMgen}
\frac{\Omega_\psi}{\Omega_{S}}=\frac{m_\psi (\eta_D+Y_{\rm FI}^{})}{\eta_Dm_{S}f(R)}\,,\qquad
\frac{\Omega_{\rm DM}}{\Omega_B}=\frac{m_\psi (\eta_D+Y_{\rm FI})+\eta_D  m_{S}f(R)}{\eta_B (1+R) m_p}\,,
\end{align}
where
\begin{equation}
f(R) \equiv
\begin{cases}
1+2R \qquad \text{if } T_D^{(S)}<T_*^{(S)} \\
1 \qquad\qquad\hspace{2mm}\text{if } T_D^{(S)}>T_*^{(S)}\,.
\end{cases}
\end{equation}

\begin{figure}[!htb]
	\centering
	\includegraphics[width=0.7\textwidth]{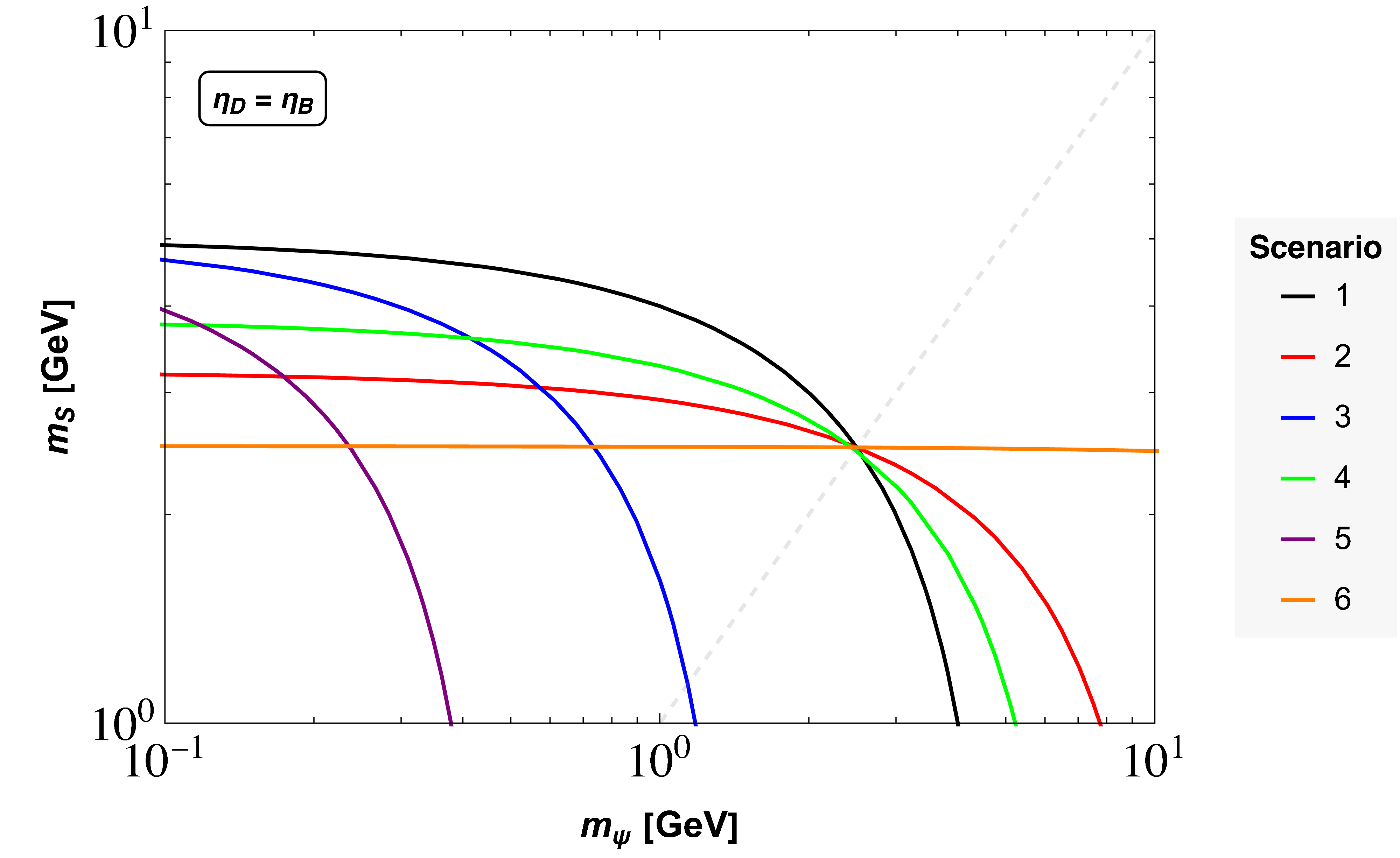}
	\caption{Contours of DM relic abundance in the $m_\psi$ - $m_S$ plane corresponding to $\eta_D = \eta_B$ for the different scenarios discussed above. We fix the values of the yukawa couplings $(y_\phi,|y_S|)$ for each scenario: \textit{Scenario 1}: $(10^{-12},10^{-3})$, \textit{Scenario 2}: $(10^{-12},5\times10^{-2})$, \textit{Scenario 3}: $(10^{-10},10^{-3})$, \textit{Scenario 4}: $(6.4\times10^{-12},10^{-1})$, \textit{Scenario 5}: $(2\times10^{-10},2\times10^{-4})$, \textit{Scenario 6}: $(10^{-13},5\times10^{-2})$. The gray dashed line corresponds to $m_\psi =m_S$.}  \label{fig:all}
\end{figure} 

In Fig.~\ref{fig:all} we show contours of correct relic abundance in the $m_S$ versus $m_\psi$ plane for the different scenarios. For definiteness, we consider the following benchmark point for all our plots: $m_\chi=3.5$ TeV, $m_{Z_D}=500$ GeV, $g_D=0.5$, $M_{N_1}=10^{11}$ GeV and $m_\nu=0.05$ eV. We take $\eta_D=\eta_B$ and for each scenario we fix an appropriate value for the Yukawa couplings $|y_S|$ and $y_\phi$ (Scenarios 1, 2  and 5 are also shown in Figs.~\ref{fig:scenario1} and \ref{fig:scenario3}, respectively). DM could be mainly composed by $\psi$, with $m_\psi$ in between hundreds of MeV and tens of GeV, mainly by the scalar $S$ with $m_S\sim\text{GeV}$, or by a combination of them (both GeV-ish). For fixed $R$, $\psi$ DM is heavier when asymmetric (Scenario 1) and lighter when symmetric (Scenario 5). For fixed $r_\psi$, $\psi$ gets heavier as $R$ gets larger. On the contrary, $S$ DM gets lighter while $R$ grows. The minimal value of $m_S$ is fixed by Eq.~\eqref{eq:mass6}, corresponding to Scenario 6. Notice the presence of a four-fold degeneracy between scenarios $1, 2, 4, 6$. This corresponds to the case $m_S=m_\psi$, in which the mass of both the DM particles is fixed by Eq.~\eqref{eq:mass6} (the relation is exact for Scenarios 1, 2 and 6, while for Scenario 4 it is approximately valid if $Y_{\rm FI}<\eta_D$). Choosing another value for $\eta_D$ leads to a rescaling of DM masses by a factor $\eta_B/\eta_D$.

\begin{figure}[!htb]
	\centering
	\includegraphics[width=1\textwidth]{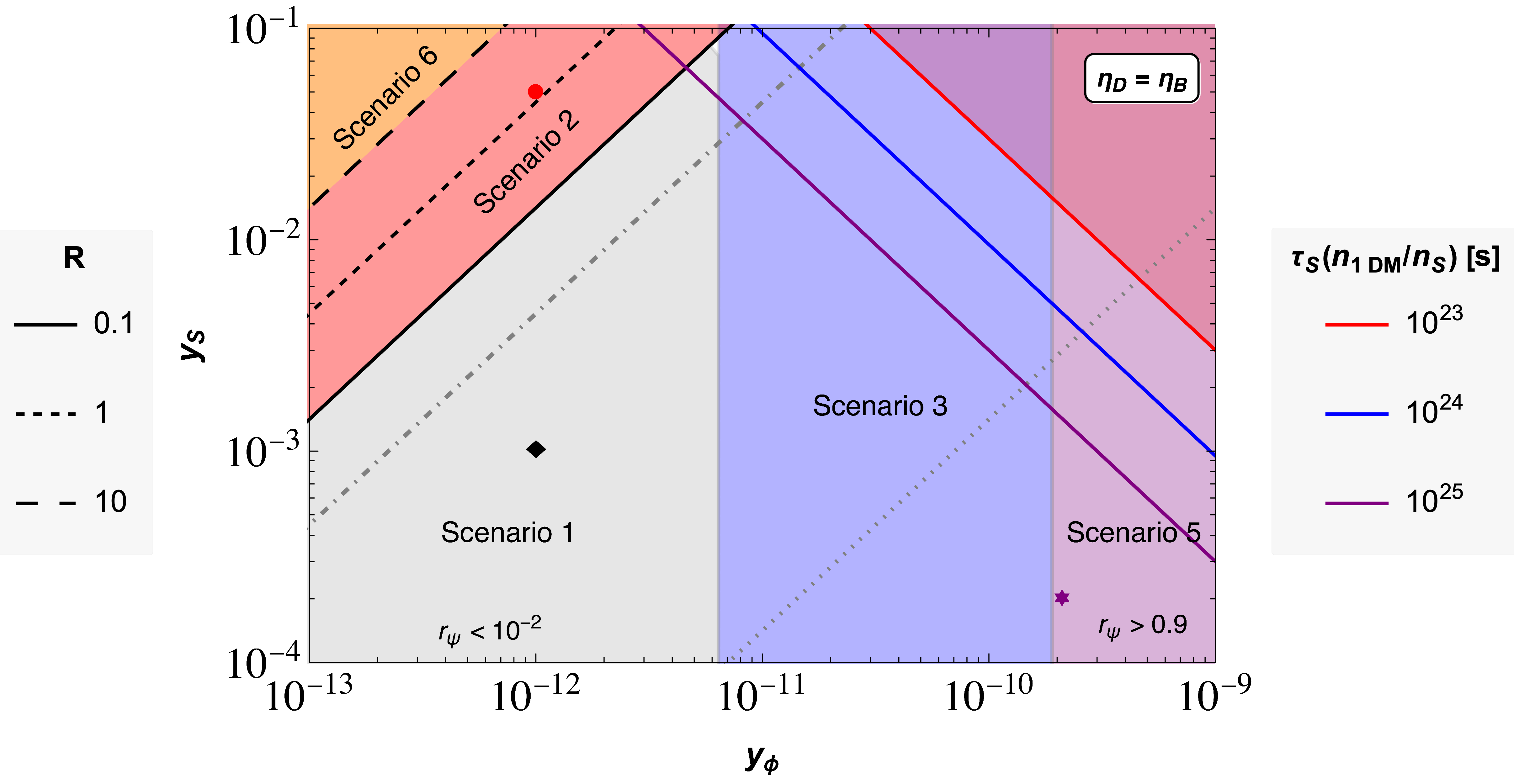}
	\caption{Parameter space of the scenarios discussed above. We fix $m_\chi=3.5$ TeV, $m_{Z_D}=500$ GeV, $g_D=0.5$, $M_{N_1}=10^{11}$ GeV, $m_\nu=0.05$ and $\eta_D=\eta_B$. Scenario 4 is not visible in the plot but would appear in the intersection between Scenario 2 (red region) and Scenario 3 (blue region). Depending on the value of $m_S$, a portion of the region in Scenario 2 (3) could convert into the Mixed Scenario 1-2 (3-4); however, for $m_S\gtrsim$ GeV, this requires $|y_S|\gtrsim0.1$. The gray dot-dashed (dotted) line corresponds to $R=1$ for $M_{N_1}= 10^{9}$ $(10^4)$ GeV. Above the red (blue) [purple] line, the heavier between $S$ and $\psi$, for \textit{fixed mass} $\max(m_S, m_\psi)=3$ GeV, has a lifetime, re-scaled by its relative number density, of $10^{23}\,(10^{24})\, [10^{25}]$ s. Therefore, the region above the red line is excluded, see Section~\ref{sec:pheno}.} \label{fig:summary}
\end{figure} 

The main results of the paper are provided in Table~\ref{tab:sum} and Fig.~\ref{fig:summary} where we show the requirements on $y_\phi$ and $R$ for each scenario, and show the allowed parameter space in the plane $|y_S|$ versus $y_\phi$, respectively. We restrict our analysis to the region $10^{-13}\lesssim y_\phi\lesssim 5\times10^{-7}$ (no thermalisation of $\psi$ and BBN bound on $y_\phi$ fulfilled for $R<1$). The Yukawa $|y_S|$ is large enough to generate a sizeable dark asymmetry (and respects the BBN bound in the region in which $R>1$, i.e., $|y_S|\gtrsim10^{-3}$). The gray dot-dashed (dotted) line corresponds to $R=1$ for $M_{N_1}= 10^{9}$ $(10^4)$ GeV. Clearly, as $N_1$ gets lighter, $\chi$ preferably decays into $S^\dagger$. Notice that in the plot the masses of the DM particles are not fixed, but at every point there are always some values of the latter for which the correct DM relic abundance is reproduced.

\section{Phenomenological signals}\label{sec:pheno}

In principle, the models considered in this work may be difficult to test and disentangle in their current version, because of several reasons:
\begin{itemize}
\item[\emph{i)}] Freeze-in scenarios invoke very small couplings, $g_X\ll 1, y_\phi \ll 1$; 
\item[\emph{ii)}] The considered scale of $B-L$ breaking is very large, $v_{\rm B-L} \gg v_{\rm EW}$, so collider searches are not an option; 
\item[\emph{iii)}] Asymmetric DM yields suppressed indirect detection signals in general. Moreover, in our scenarios, the symmetric component is typically erased into the dark sector, via $\bar{\chi} \chi \rightarrow Z_D Z_D$ and $S^\dagger S \rightarrow \varphi \varphi$, see Fig.~\ref{fig:chiann}. 
\item[\emph{iv)}] The mixing of $\varphi$ with the Higgs, generated by $\lambda_{H\phi} H^\dagger H \phi^\dagger \phi$, was taken to be very small.
\end{itemize}

However, there are a few distinctive signals of $S$, through the usual Higgs portal, $\lambda_{HS} H^\dagger H S^\dagger S$:
\begin{itemize}

\item Direct detection in the case in which the DM is mainly composed of $S$, which currently sets the limits $\lambda_{HS}\lesssim 0.01$ \cite{Cline2013}.

\item Higgs invisible decays, for $m_S<m_h/2$, which currently sets the limits $\lambda_{HS}\lesssim 0.01$ \cite{Clarke:2013aya,Cline2013}.  In this case, it could be that $S$ was produced via Higgs decays, but it did not constitute a dominant part of the DM. 

\item Similarly, indirect detection signals from annihilations are present if the final abundance is composed of $S$ and partially-asymmetric (Scenarios 2 and 4) or  symmetric (Scenario 6). In this case, on-shell $S$-annihilations into muons, pions and electrons are possible. Typically, such light thermal DM is severely constrained by its energy injection in the CMB. In particular, Scenario 6 is characterized by an enhanced indirect detection signal due to larger than usual annihilation rates. Furthermore, it is very predictive, with a DM mass of $2.5$ GeV, see Eq.~\eqref{eq:mass6}. Therefore, it may be interesting to further investigate for which values of $\lambda_{HS}$ is Scenario 6 allowed.

\end{itemize}

Scenario 6 is also interesting as the $S^\dagger$ population arising from late decays of $\chi$, results in a mixture of cold/warm DM, where $S$ particles coming from the thermal plasma, represent the cold component, while their anti-particles, coming from late decays, the warm one, with a possible impact on structure formation \cite{Falkowski:2011xh}. Furthermore, the additional contribution to the asymmetric background neutrino population is maximal in this case, $\Delta Y_\nu\approx \eta_D$.

Last but not least, there is a very interesting phenomenological signal of our model: the presence of a monochromatic flux of neutrinos coming from the late decay of the heaviest of the two DM components. Let us assume $m_S>m_\psi$ (the opposite case $m_\psi>m_S$ is completely analogous). At low energies, $E\ll m_\chi\ll M_{N_1}$, the decay $S\to \bar{\psi}+\nu_L$ is generated by the dimension-6 operator
\begin{equation}
\mathcal{O}_6=\bar{L}\tilde H S\phi^\dagger\psi\,.
\end{equation} 
This operator arises by first integrating the right-handed neutrino field $N_R$, which generates the interaction given in Eq.~\eqref{eq:OD}, and then at the lower scale the fermion $\chi$, giving rise to the interaction \footnote{Operator $\mathcal{O}_6$ may also be generated by a UV completion of our model at scales above $v_{B-L}$. This produces additional constraints, weaker than Eq.~\eqref{meq:mSlimit1}. We discuss them in Appendix~\ref{sec:stability}.}
\begin{equation}
\frac{y_Sy_\phi y_\nu}{M_{N_1}m_\chi}\mathcal{O}_6\,.
\end{equation}
Once $H$ and $\phi$ acquire vevs, the decay $S\to \bar{\psi} +\nu_L$ is generated. 
The decay width reads
\begin{equation}
\Gamma(S\to \bar{\psi}+\nu_L)\approx \frac{|y_S|^2y_\phi^2 m_S}{32\pi}\left(\frac{v_\phi}{m_\chi}\right)^2 \left(\frac{m_\nu}{M_{N_1}}\right)\left(1-\frac{m_\psi^2}{m_S^2}\right)\,.
\end{equation}
To guarantee that both $S$ and $\psi$ are cosmologically stable and contribute to the DM abundance, the lifetime of the heavier particle needs to be larger than the age of the Universe, $\tau_S>t_U\sim 4\times 10^{17}$ s.
However, there are stronger constraints if the decay products include an active neutrino, $\tau_S\gtrsim10^{23}$ s (for a GeV-ish single-component DM) \cite{Palomares-Ruiz:2007egs,Bell:2010fk}. If $S$ decays at late times, it leads to a very distinctive signature: a neutrino line peaked at $m_S/2\sim \mathcal{O}(\text{GeV})$. Therefore, depending on the relative abundance of $\psi$ and $S$, some region of the parameter space could be excluded, see Refs.~\cite{Garcia-Cely:2017oco,ElAisati:2017ppn,Coy:2021sse}. To quantify this, we compare the experimental bound with the re-scaled lifetime of the particle, $\tau_S^{\rm re-sc}=\tau_S\,(n_{\rm 1DM}/n_{\rm S})>10^{23}$ s, where $n_{\rm 1DM}$ ($n_S$) is the single-component DM ($S$) number density. This constraint gives a condition on the parameters,
\begin{equation}\label{meq:mSlimit1}
\begin{split}
y_\phi |y_S|
&\lesssim 3\times10^{-12}\left(\frac{3\text{ GeV}}{m_S}\right)^{1/2}\\
&\left(\frac{700\text{GeV}}{v_\phi}\right)\left(\frac{m_\chi}{3.5\text{TeV}}\right)\left(\frac{0.05\text{eV}}{m_\nu}\right)^{1/2}\left(\frac{M_{N_1}}{10^{11}\text{GeV}}\right)^{1/2}\left(\frac{\Omega_{\rm DM}}{\Omega_S(y_\phi,|y_S|,m_S,\eta_D)}\right)^{1/2}\,,
\end{split}
\end{equation}
valid if $m_S>m_\psi$. In general the condition is not linear in the parameters. If $\Omega_S/\Omega_{\rm DM}\ll1$, the constraint is extremely weak: DM is made only by $\psi$ and there are no decays. On the contrary, if $\Omega_S/\Omega_{\rm DM}\sim\mathcal{O}(1)$, the condition becomes linear in the Yukawas.
In the opposite regime, $m_\psi>m_S$, there is an equivalent condition to that of Eq.~\eqref{meq:mSlimit1}, with $m_S\to m_\psi$ and $\Omega_S\to\Omega_\psi$. If the two states are almost degenerate $m_S\simeq m_\psi$ there is a strong phase space suppression so that the constraint gets weaker.

Notice that, even taking into account that the lifetime is re-scaled by the relative number density, this constraint is much stronger than the one coming from $\tau_S>t_U$, so that it automatically guarantees the stability of both $S$ and $\psi$. 

The neutrino line emerging from $S\to \bar{\psi}\nu_L$ ($\psi\to S^\dagger\nu_L$) decay is the main smoking gun of our scenarios. As an illustrative example, in Fig.~\ref{fig:summary} we show three different lines in red, blue and purple corresponding respectively to $\tau_S^{\rm re-sc}=10^{23},10^{24},10^{25}$ s, for fixed $\max(m_S,m_\psi)=3$ GeV. For this choice of mass, as we can see in Fig.~\ref{fig:all}, $\Omega_{\rm S}/\Omega_{\rm DM}\sim\Omega_{\rm \psi}/\Omega_{\rm DM}\sim\mathcal{O}(1)$;  therefore, Eq.~\eqref{meq:mSlimit1} gives a simple linear constraint on the Yukawas and the DM masses. The corresponding region above the red line is excluded. In the region between the red and the purple lines the signal is close to the experimental sensitivity and could lead to the observation of a neutrino line in existing or near-future neutrino telescopes (see Refs.~\cite{SuperK:2003,SuperK:2015,SuperK:2016,Borexino:2011,KamLAND:2012,Palomares-Ruiz:2020ytu}).  From the figure we deduce that Scenarios 3 and 5 are therefore the most testable ones. Notice that, for different choices of $\max(m_S,m_\psi)$, the picture can be more complicated because Eq.~\eqref{meq:mSlimit1} becomes non-linear. By performing a numerical scan over the relevant Yukawa and DM mass ranges, we have checked  that, as expected, there is always an excluded region in the upper right corner of Fig.~\ref{fig:all}, which corresponds to $y_\phi\gtrsim10^{-10}$, $|y_S|\gtrsim0.03$.

\section{A low-energy variant: The inverse seesaw} \label{sec:ISS}

So far, we have assumed that active neutrinos get masses through the Type-I seesaw at a very high scale. In this framework, the lepton and dark asymmetries are generated in a similar way as in high-scale leptogenesis. In the following, we discuss the possibility to embed our models into a low-scale leptogenesis scenario, trying to lower the scale of $B-L$ of the scenarios considered in this work. This is an alternative path, which may yield interesting phenomenological signals.
In this case, there can be DM interactions with the SM, mediated by $Z_{B-L}$. Apart from direct and indirect detection signals, $\chi$ may be produced at colliders via $\bar{q}q\to Z_{B-L}\to\bar{\chi}\chi$ and then decay, $\chi\to\psi\varphi$ or $\chi \rightarrow S^\dagger \nu$, yielding missing energy.\footnote{Notice that there is no $Z_{B-L}S^\dagger S$ vertex in the Lagrangian.} For $R\lesssim 1$, if the mixing of $\varphi$ with the Higgs is in the correct range, displaced vertices may be produced from $\varphi \rightarrow \rm SM\,\, SM$. Searches for long-lived particles are very active, with lots of experiments running or designed for the following years \cite{MATHUSLA:2019,FASER:2019,SHIP:2015}.  

It has been shown that leptogenesis at a low scale is possible, for example via resonant leptogenesis, with $\mathcal{O}(10{\rm~TeV})$ right-handed neutrino masses  \cite{Pilaftsis:2003gt,Klaric:2020phc,Hugle:2018qbw}. Another possibility is to adopt an inverse see-saw (ISS) mechanism to give mass to neutrinos~\cite{Mohapatra:1986aw,Mohapatra:1986bd}. Next we consider this option within a $B-L$ set-up, see also Refs.~~\cite{Kajiyama:2012xg,Abada:2021yot,Panda:2022kbn,Hirsch:2009ra,Garayoa:2006xs} and the review \cite{Cai:2017jrq}. We can modify our model by replacing the scalar $\sigma$ with two new fields: a scalar $\sigma'$ and three copies of a fermion $S_L$, with the quantum numbers outlined in Table \ref{tab:inverse}.

\begin{table}[!htb]
	\centering
	\begin{tabular}{c c c c c}
		\hline
		Field & Spin & $U(1)_{B-L}$ & $U(1)_D$ & $U(1)_X$ \\
		\hline
		$S_L$ & 1/2 & 0 & 0 & 0   \\
		$\sigma'$ & 0& +1 & 0 & 0 \\
		\hline
	\end{tabular}
	\caption{Quantum numbers of the new states in the inverse seesaw variant.\label{tab:inverse}}
\end{table} 
The Lagrangian is the same as Eq.~\eqref{eq:Lagnew} (excluding the terms involving $\sigma$), with the addition of 
\begin{equation}
\mathcal{L}_{\rm ISS}=\overline{S_L}i\slashed\partial S_L- \sigma' \overline{S_L} y_{\sigma'} N_R- \frac{1}{2}\overline{S}_L \mu S^c_L+ {\rm H.c.}\,,
\end{equation}
where $\mu$ is a $3 \times 3$ complex symmetric matrix which can be taken to be real and diagonal without loss of generality, and $y_{\sigma'}$ is a $3 \times 3$ complex matrix. If the new scalar takes a vev, $\med{\sigma'}=v_{B-L}$, $B-L$ is spontaneously broken and $S_L$ and $N_R$ form a pseudo-Dirac pair, with mass $M_D=y_{\sigma'}\, v_{B-L}$. In this case, neutrinos get a mass through the inverse see-saw mechanism. We focus on the range $M_D>m_D\gg \mu$. The mass of the active neutrinos is given by
\begin{equation}
m_\nu \simeq  m_D\,M_D^{-1}\, \mu\, (M_D^{-1})^T m^T_D\,,
\end{equation}
so that light neutrino masses may be reproduced with small values of $M_D\sim v_{B-L}\sim \mathcal{O}$ (TeV) for small enough values of $\mu$.

Both low-scale variants result into a massive gauge boson, $Z_{B-L}$, with a mass in the $5-10$ TeV range, allowed by experimental constraints \cite{Escudero:2018fwn}. In this case, the larger $B-L$ gauge interactions allow to erase the symmetric $\chi$ population. Therefore, the $U(1)_D$ gauge interactions are not needed, and in the limit $g_D\to0$, $U(1)_D$ acts as a global symmetry.\footnote{In the limit $g_D\to0$, the unbroken $U(1)_{X+D}$ symmetry is now global. The would-be Goldstone boson of the broken $U(1)_{X-D}$ is now eaten by $A'$, which becomes massive. $A'$ is light because its mass is proportional to the gauge coupling $g_X$, $m_{A'}\sim g_X v_\phi\ll v_\phi$ (recall that $g_X$ is tiny by assumption to avoid $\psi$ thermalisation), and it does not thermalise because it has only gauge interactions driven by $g_X$. $A'$ couples to the $B-L$ current proportionally to $(v_\phi^2/v_{B-L}^2)g_X$. The interactions of a massive gauge boson are less constrained than those of a massless one. However, as now $v_\phi\lesssim v_{B-L}$, one obtains a constraint of the order of $g_X\lesssim 10^{-12}$.} Notice that the presence of a global $U(1)_D$ is still crucial to stabilise the DM particles and forbid dangerous operators, such as $\bar{\psi}\phi N_R$, which would generate $\psi-N_R$ mixing. However, $U(1)_D$ could also be replaced by a $Z_2$ symmetry, under which all fields but $\chi$, $S$ and $\phi$ are even. In such a case, the symmetry stabilising the DM particles is a $Z'_2$, coming from the combination of the broken $U(1)_X$ and the original $Z_2$.

\begin{figure}[!htb]\label{fig:pslow}
\centering
	\includegraphics[width=0.6\textwidth]{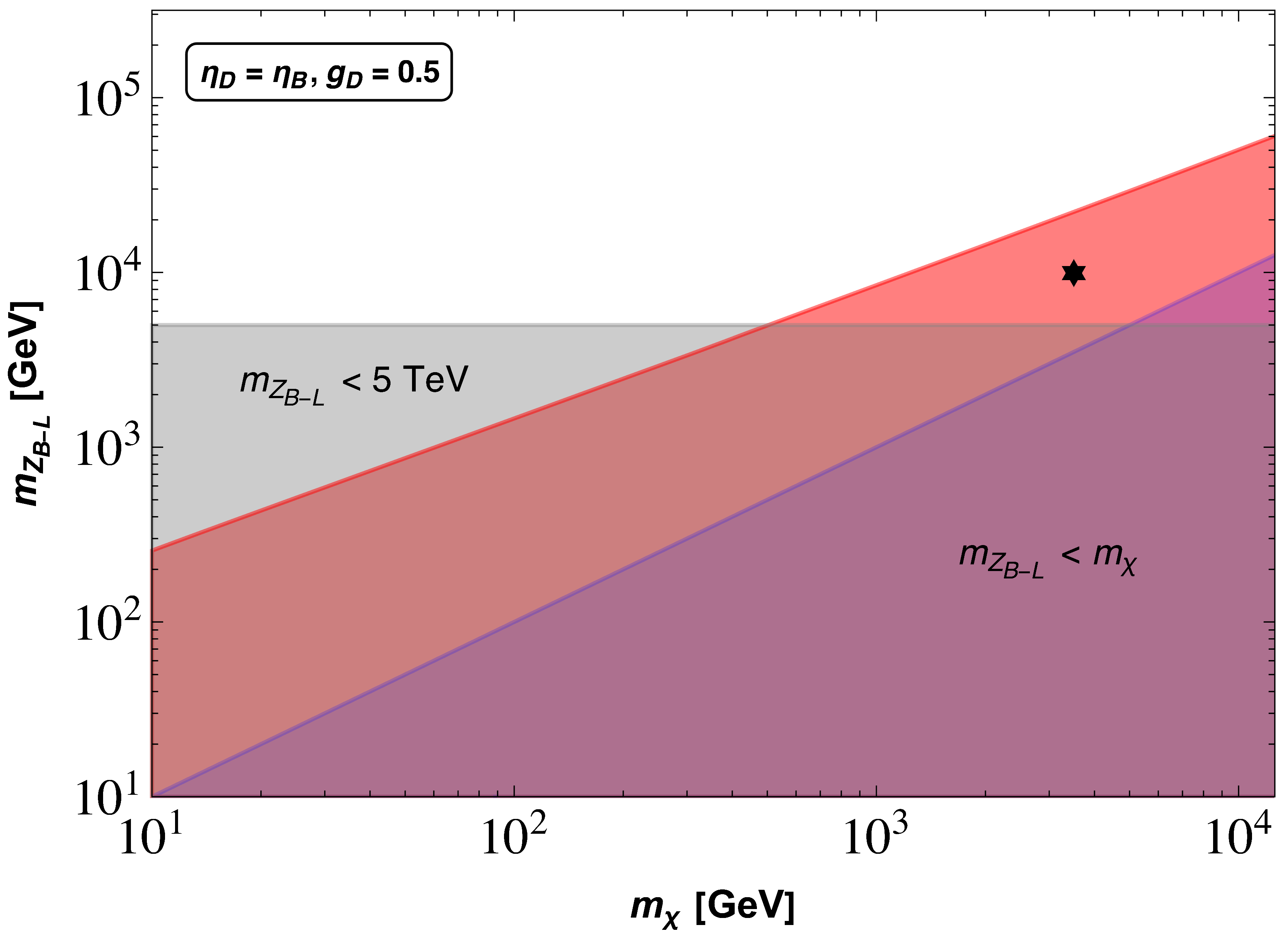}
	\caption{Parameter space for the low-scale variant with the gauge coupling fixed at 0.5 and $\eta_D = \eta_B$. In the red region the fractional asymmetry $r_\chi <10^{-2}$. The gray region is excluded because $Z_{B-L}$ is too light whereas in the purple region $Z_{B-L}$ is lighter than $\chi$. Since we expect $m_{Z_{B-L}}\gtrsim m_{N_1}$, in the purple region $N_1\to\chi S$ decays are forbidden. The star indicates the benchmark point $\{m_\chi,m_{Z_{B-L}}\}=\{3.5\text{ TeV}, 10\text{ TeV}\}$.}
\end{figure}

Thanks to the low scale, the annihilations $\bar{\chi}\chi\to Z_{B-L}\to \bar{q}q\, (\bar{l}l)$ are strong enough to erase the symmetric population of $\chi$ and make the model more testable. 
The cross section for the $\bar{\chi}\chi\to Z_{B-L}\to \bar{f}f $ process (assuming we are far enough from the resonance, i.e., $4m_\chi^2-m_{Z_{B-L}}^2\gg \Gamma_{Z_{B-L}} m_{Z_{B-L}}$), reads
\begin{equation}
\sigma v(\bar{\chi}\chi\to \bar{f}f)=\frac{N_C(f)g_{B-L}^4q_{B-L}^2(f)}{2\pi}\,\sqrt{1-\frac{m_f^2}{m_\chi^2}}\,\frac{2m_\chi^2+m_f^2}{(4m_\chi^2-m_{Z_{B-L}}^2)^2}\,,
\end{equation}
where $N_C (f)=3\,(1)$ and $q_{B-L}(f)=1/3\,(-1)$ for quarks (leptons). The parameter space where these annihilations lead to $r_\chi < 10^{-2}$ is shown in Fig.~\ref{fig:pslow}, with $g_D = 0.5$ and $\eta_D = \eta_B$. 

The different scenarios for DM are similar to the ones studied in the previous sections. However, the decay $\chi\to S^\dagger\nu_L$ is now enhanced (with respect to the previous case) by the larger value of $(m_\nu/M_{N_1})$.
Therefore, we need smaller values of $|y_S|$ to realise a scenario with $R\ll1$. However, too small $|y_S|$ may be problematic for the generation of the dark asymmetry. A precise lower bound is not well established and could be computed in future works. Assuming $|y_S|\gtrsim10^{-4}$, the most natural scenario  is that the decays of $\chi$ into $S^\dagger$ occur while the latter are still in equilibrium, at $T_D^{(S)}>T_*^{(S)}$, (partially) washing-out the asymmetry $\eta_S$.
Therefore, if $y_\phi$ is small enough such that $R>10$, $\chi$ mostly decay into $S^\dagger$ and $\eta_S$ gets completely erased. Subsequently $S$ undergoes a standard symmetric freeze-out and its abundance is determined by the annihilation cross section $\sigma v_{S^\dagger S}$. 

If $y_\phi$ is larger, other scenarios can be realised, such as Scenarios 3, Mixed 3-4 or 5. Eventually, if both $y_\phi>2\times 10^{-10}\sqrt{\eta_D/\eta_B}$ and $R>1$, a possibility not realised when $v_{B-L}\gg$ TeV, a new scenario appears: $\psi$ is symmetric and is produced mainly by freeze-in with abundance $Y_\psi^+\simeq Y_\psi^-=Y_{\rm FI}/(2(1+R))$ while the decays into $S^\dagger$ partially washout $\eta_S$, leaving an asymmetric abundance of $S$, $Y_S^+=\eta_D/(1+R)\gg Y_S^-$. This is basically a generalisation of Scenario 5 with all the abundances rescaled by a $1+R$ factor. Notice that, in any case, in order to have multicomponent DM, $S$ must be light enough to satisfy the neutrino constraints, corresponding to the decay $S\to \psi\nu_L$ discussed in Section~\ref{sec:pheno}, which get stronger the smaller the $v_{B-L}$ scale. Interestingly, in this case regions of smaller $y_\phi$ and $|y_S|$ could be probed.

\section{Conclusions}\label{sec:conclusions}

The dark sector may be very rich. If this is the case, there is a plethora of possibilities regarding the number of stable particles, their nature, and their production mechanisms. In this work we classify and analyse the different options by adopting a cogenesis model that simultaneously explains neutrino masses, the baryon asymmetry and the DM relic abundance. While neutrino masses and the baryon asymmetry are produced via the standard Type-I seesaw mechanism and leptogenesis (with some extra contributions), respectively, we find that in such a framework there is a variety of viable scenarios for explaining the nature and abundance of dark matter.

Once the decays of right handed neutrinos into the visible and dark sectors generate the asymmetries, some dark sector particles undergo asymmetric freeze-out and others are produced via freeze-in. The model has two potential DM candidates, and we focus on the parameter space where both particles are stable. However, whether they both contribute to the DM abundance similarly (two-component DM) or one has a negligible contribution (one-component DM) depends on their asymptotic asymmetries, where the late decays play an important role. Such decays may significantly populate the asymmetric or symmetric component at later times, thereby restoring annihilations, which may lead to enhanced signals in DM indirect detection. In this case, even though the DM is symmetric at the end, its abundance is still set by the asymmetry, and is thus independent of the annihilation rate, contrary to the usual WIMP scenario. 

We have analysed the range of model parameters that control the contribution of each component to the DM abundance, and outline the possible scenarios, classified according to the nature and production mechanism of each particle. We consider DM masses in the GeV ballpark and dark asymmetries similar to the baryonic one; however, the set-up can easily accommodate lighter (heavier) DM for a larger (smaller) dark sector asymmetry if the branching ratio of right handed neutrinos into the dark sector is smaller (larger). We have found that one of the main distinctive signatures is a neutrino line from $S$ (or $\psi$) decays. This would constitute a smoking gun of our model, within reach of existing or near-future neutrino telescopes for a significant region of the parameter space of some of the scenarios.

We conclude that having an initial asymmetry in the dark sector does not necessarily predict completely asymmetric dark matter, with its mass constrained by the dark asymmetry. In extended models, it allows the DM component to be partially asymmetric or symmetric, leading to more flexibility regarding the DM mass as well as the phenomenological implications. Finally, although in this work we focused and extended the cogenesis scenario, which relates neutrino physics and dark matter, it would be interesting to consider other frameworks in which the different possibilities outlined here may also be present.

\vspace{1cm}

\acknowledgments

This work is supported by the MICIN/AEI (10.13039/501100011033) grants PID2020-113334GB-I00 and PID2020-113644GB-I00. GL is supported by the European project H2020-MSCA-ITN-2019/860881-HIDDeN. JHG and DV are supported by the “Generalitat Valenciana” through the GenT Excellence Program (CIDEGENT/2020/020). 

\appendix

\section{Details of the different scenarios}\label{app:scenarios}

\begin{table}[!htb]
		\centering
		\begin{tabular}{p{0.04\textwidth}>{\centering}p{0.27\textwidth} >{\centering}p{0.22\textwidth} >{\centering}p{0.24\textwidth} >{\centering\arraybackslash}p{0.14\textwidth}}
			\hline
			\textbf{Sc.}  & $\psi$  & $S$ & $\Omega_{\rm DM}/\Omega_B$ & $\Omega_S/\Omega_\psi$ \\
			\hline
			\thead{\textbf{1}} & \thead{Asymmetric \\ LD $\chi\to\psi\varphi$ \\ $Y_\psi^+ =\eta_D$ \\ $Y_{\psi}^-\ll Y_\psi^+$}  & \thead{Asymmetric\\FO $S^\dagger S\to \varphi\varphi$\\$Y_{S}^+=\eta_D$\\$Y_{S}^- \ll Y_{S}^+$} & $\frac{\eta_D}{\eta_B}\frac{m_\psi+m_{S}}{ m_p}$ & $\frac{m_{\psi}}{m_{S}}$ \\		
			\hline
			\thead{\color{red}{\textbf{2}}}	 & \thead{Asymmetric\\LD $\chi\to\psi\varphi$\\$Y_\psi^+=\eta_D/(1+R)$\\$Y_{{\psi}}^-\ll Y_\psi^+$}  & \thead{Partially asymmetric\\FO $S^\dagger S\to \varphi\varphi$\\+ LD $\chi\to S^\dagger \nu_L$\\$Y_{S}^{+}=\eta_D$\\$Y_{S}^-={\eta_DR}/{(1+R)}$} &$\frac{\eta_D}{\eta_B}\frac{m_\psi+(1+2R)m_{S}}{(1+R) m_p}$  & $\frac{m_{\psi}}{m_{S}(1+2R)}$  \\				 
			\hline			
			\thead{\textbf{1-2}} & \thead{Asymmetric\\LD $\chi\to\psi\varphi$\\$Y_\psi^+=\eta_D/(1+R)$\\$Y_{{\psi}}^-\ll Y_\psi^+$}  &  \thead{Asymmetric\\FO $S^\dagger S\to \varphi\varphi$\\+ LD $\chi\to S^\dagger \nu_L$\\$Y_{S}^{+}=\eta_D/(1+R)$\\$Y_{S}^-\ll Y_{S}^+$ } & $\frac{\eta_D}{\eta_B}\frac{m_\psi+m_{S}}{(1+R) m_p}$ & $\frac{m_{\psi}}{m_{S}}$\\				 
			\hline				
			\thead{\color{blue}{\textbf{3}}}& \thead{Partially asymmetric\\FI + LD $\chi\to\psi\varphi$\\$Y_{\psi}^+= Y_{\rm FI}/2+\eta_D$\\$Y_{{\psi}}^-= Y_{\rm FI}/2$} & \thead{Asymmetric\\FO $S^\dagger S\to \varphi\varphi$\\$Y_{S}^+=\eta_D$\\$Y_{S}^-\ll Y_{S}^+$} & $\frac{m_\psi (\eta_D+Y_{\rm FI})+\eta_D m_{S}}{\eta_B m_p}$& $\frac{m_\psi (\eta_D+Y_{\rm FI}^{})}{m_{S}\eta_D}$ \\		
			\hline			
			 \thead{\color{green!50!black}{\textbf{4}}}&  
			 \thead{Partially Asymmetric\\FI +LD $\chi\to\psi\varphi$\\$Y_{\psi}^+= ({Y_{\rm FI}}/{2}+\eta_D)/(1+R)$\\$Y_{{\psi}}^-= Y_{\rm FI}/(2(1+R))$} & \thead{Partially Asymmetric\\FO $S^\dagger S\to \varphi\varphi$\\+ LD $\chi\to S^\dagger\nu_L$\\$Y_{S}^+=\eta_D$\\$Y_{S}^-=\eta_D R/(1+R)$} & $\frac{m_\psi (\eta_D+Y_{\rm FI})+\eta_D (1+2R) m_{S}}{\eta_B (1+R) m_p}$ & $\frac{m_\psi (\eta_D+Y_{\rm FI}^{})}{m_{S}\eta_D(1+2R)}$\\			 	
			\hline			
			\thead{\textbf{3-4}} &  \thead{Partially Asymmetric\\FI +LD $\chi\to\psi\varphi$\\$Y_{\psi}^+= ({Y_{\rm FI}}/{2}+\eta_D)/(1+R)$\\$Y_{{\psi}}^-= Y_{\rm FI}/(2(1+R))$} &  \thead{Asymmetric\\FO $S^\dagger S\to \varphi\varphi$\\+ LD $\chi\to S^\dagger\nu_L$\\$Y_{S}^+=\eta_D/(1+R)$\\$Y_{S}^-\ll Y_S^+$} &$\frac{m_\psi (\eta_D+Y_{\rm FI})+\eta_D  m_{S}}{\eta_B (1+R) m_p}$ & $\frac{m_\psi (\eta_D+Y_{\rm FI}^{})}{m_{S}\eta_D}$\\		    		
			\hline				
			\thead{\color{purple!50!black}{\textbf{5}}} &	\thead{Symmetric\\ FI $\chi\to\psi\varphi$\\$Y_{\psi}^+= {Y_{\rm FI}}/{2}+\eta_D\simeq  {Y_{\rm FI}}/{2} $\\$Y_{{\psi}}^-= Y_{\rm FI}/2$} & \thead{Asymmetric\\FO $S^\dagger S\to \varphi\varphi$\\$Y_{S}^+=\eta_D$\\$Y_{S}^-\ll Y_{S}^+$} & $\frac{\eta_D}{\eta_B}\frac{m_\psi (Y_{\rm FI}/\eta_D)+m_{S}}{m_p}$ & $\frac{m_\psi Y_{\rm FI}^{}}{m_{S}\eta_D}$\\			 		
			\hline
			\thead{\color{orange!90!black}{\textbf{6}}} &	\thead{Negligible production}  & \thead{Symmetric\\FO $S^\dagger S\to \varphi\varphi$\\+ LD $\chi\to S^\dagger \nu_L$\\$Y_{S}^+=\eta_D$\\$Y_{S}^-=\eta_D$} & $<1$ &$\frac{\eta_D}{\eta_B}\frac{2m_{S}}{m_p}$ \\			
			\hline
		\end{tabular}
		\caption{\label{tab:scesum} Classification of scenarios in our model on the basis of dominant production mechanism and asymptotic nature of both dark matter components, $\psi$ and $S$.}
	\end{table}

\subsection*{Scenario 1}

In this scenario, both components are asymmetric as we take $y_\phi\lesssim6\times10^{-12}\sqrt{\eta_D/\eta_B}$ and $R\ll1$, so that the freeze-in population of $\psi$ from early decays is negligible, while the late decays are dominant, resulting in asymmetric $\psi$ with abundance  $Y_\psi^+=\eta_\psi=\eta_D$, whereas $S$ freezes-out once all the symmetric population has annihilated. The population of $S^\dagger$ produced by the late decays of $\chi$ is negligible. The abundance is determined by its asymmetry, i.e $Y_{S}^+=\eta_{S}=\eta_D$. Hence, both the components have individual abundances $\eta_D$ and we have
\begin{align}\label{eq:DMabb}
\frac{\Omega_{\psi}}{\Omega_S}= \frac{m_{\psi}}{m_S}\,,\qquad
\frac{\Omega_{\rm DM}}{\Omega_B}\simeq 5 =\frac{\eta_D(m_\psi+m_{S})}{m_p\eta_B}\,.
\end{align}
In Fig.~{\ref{fig:scenario1} (black lines), we show the region of the parameter space in which the correct DM relic abundance is reproduced in the plane $m_S$ versus $m_\psi$, for two values of the dark asymmetry $\eta_D$. $\Omega_{\psi}/\Omega_S$ increases when the curves are followed clockwise. We can see that for $\eta_D \simeq 0.1\, (1)\,\eta_B$, dashed (solid) black line, we have  $m_\psi \simeq m_{S}\simeq 30\, (3)$ GeV if both species contribute similarly. Alternatively, we can push the mass of $S$ down to GeV in such a way that its contribution to DM abundance is negligible and reproduce the relic abundance for $m_\psi \simeq 50\,(5)$ GeV for $\eta_D \simeq 0.1\, (1)\,\eta_B$, or vice versa.
	\begin{figure}[!htb]
		\centering
		\includegraphics[width=0.6\textwidth]{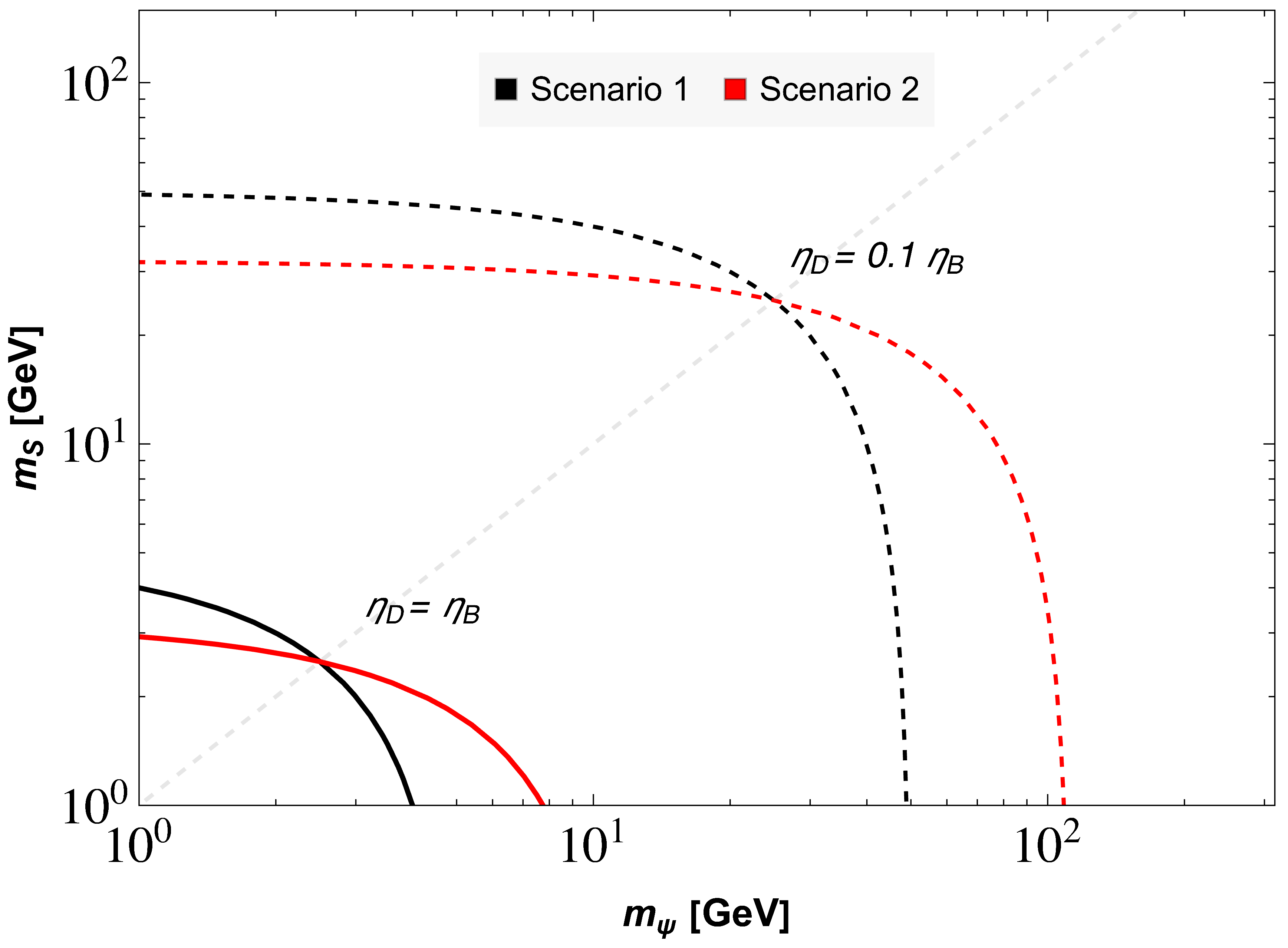}
		\caption{\label{fig:scenario1} Values of $m_\psi$ and $m_{S}$ for which the correct relic abundance can be reproduced for Scenario 1 (black lines) and Scenario 2 (red lines). We show  $\eta_D \simeq (0.1)\eta_B$ using solid (dashed) lines. The relative abundance of $\psi$ with respect to $S$ increases when the curves are followed clockwise. The gray dashed line corresponds to $m_\psi = m_S$.  We fix $m_\chi=3.5$ TeV, $m_{Z_D}=500$ GeV, $g_D=0.5$, $M_{N_1}=10^{11}$ GeV, $m_\nu=0.05$ eV. Furthermore, we fix $y_\phi=10^{-12}$. For Scenario 1\, (2): $|y_S|=10^{-3}\, (5\times10^{-2})$, so that $R\simeq0.0005\, (1.25)$. In both cases $Y_{\rm FI}\simeq 10^{-4}\eta_B$.}
	\end{figure}
	
\subsection*{Scenario 2}
	
Similar to the previous scenario, here also we take $y_\phi\lesssim 6\times10^{-12}\sqrt{\eta_D/\eta_B}$ but $R\sim \mathcal{O}(1)$ $(\text{i.e.,}~ |y_S|\sim 0.5\times 10^{11}y_\phi)$. We also assume that $|y_S|$ is small enough so that Eq.~\eqref{eq:LDS11} is satisfied, i.e., the decays of $\chi$ to $S^\dagger$ occur when the latter has already decoupled from the thermal bath, $T_D^{(S)}<T_*^{(S)}$ and $S^\dagger S$ annihilations are not active.
	
Due to $R\sim \mathcal{O}(1)$, $\chi$ partially decays into $\psi$  and partially into $S^\dagger$ with probabilities $1/(1+R)$ and $R/(1+R)$, respectively. $\psi$ is highly asymmetric with abundance $Y_\psi^+=\eta_\psi\sim \eta_D/(1+R)$, whereas $S$ becomes partially asymmetric because a population of $S^\dagger$ is produced by $\chi\to S^\dagger\nu$ decays. Therefore, $\eta_{S}=\eta_D-R\,\eta_D/(1+R)= \eta_D/(1+R)$,and its total abundance is determined by the sum $Y_{S}^+ +Y_{S}^- =(1+2R)\eta_D/(1+R)$. The contribution to the DM abundance becomes
	\begin{align}
	\frac{\Omega_{\psi}}{\Omega_S}\approx \frac{m_{\psi}}{m_S}\frac{1}{(1+2R)}\,,\qquad
	\frac{\Omega_{\rm DM}}{\Omega_B}\simeq 5=\frac{\eta_D(m_\psi+(1+2R)m_{S})}{(1+R)m_p\eta_B}\,.
	\end{align}
	As discussed earlier, this scenario leads to an enhanced background of an asymmetric neutrino population. In the limit $R=0$, all $\chi$ decay into $\psi$ and we recover Eq.~\eqref{eq:DMabb}. For $R=1$ the relative $\psi/S$ abundance is $m_\psi/(3m_S)$. The factor 3 comes from the fact that while we only have $\psi$ and no $\bar{\psi}$, we have both $S$ and $S^\dagger$ with $Y_\psi^+ =Y_{S}^- =\eta_D/2$ and $Y_{S}^+ =\eta_D$. The allowed parameter space is shown in Fig.~\ref{fig:scenario1} by red lines. In this case, the shape of the curves is not symmetric as in Scenario 1; in the limiting case where $\psi\, (S)$ dominates the abundance one needs $m_\psi\simeq100$ GeV ($m_S \simeq 30$ GeV)  for $\eta_D \simeq 0.1\,\eta_B$, and masses roughly one order of magnitude smaller for $\eta_D \simeq \eta_B$, as expected.
	
	\subsection*{Mixed Scenario 1-2}
	
	We consider the same range for $y_\phi$ and $R$ but we assume that $|y_S|$ is large enough to violate Eq.~\eqref{eq:LDS11}, i.e., $\chi$ decays to $S^\dagger$ while $S^\dagger S$ annihilations are still efficient, $T_D^{(S)}>T_*^{(S)}$. For $\psi$ we find the same results of the previous section. The $S^\dagger$ population produced by $\chi$ decays partially washes-out $\eta_S$, leaving an asymmetric population of $S$ with abundance $Y_S^+=\eta_D/(1+R)$, while the $S^\dagger$ population gets erased by $S^\dagger S$ annihilations. The contribution to the DM abundance is now
	\begin{align}
	\frac{\Omega_{\psi}}{\Omega_S}\simeq \frac{m_{\psi}}{m_S}\,,\qquad \frac{\Omega_{\rm DM}}{\Omega_B}\simeq 5=\frac{\eta_D(m_\psi+m_{S})}{(1+R)m_p\eta_B}\,.
	\end{align} 
	This scenario is a mixture between Scenario 1 (the nature of the DM particles is the same, both asymmetric, and the abundances are the same ones rescaled by $(1+R)$) and Scenario 2 (the ranges of $y_\phi$ and $R$ are identical and the same processes determine the final abundance).

	\subsection*{Scenario 3}
	
	For larger values of the Yukawa, $6\times10^{-12}\sqrt{\eta_D/\eta_B}\lesssim y_\phi\lesssim 2\times10^{-10}\sqrt{\eta_D/\eta_B}$, the freeze-in contribution to $\psi$ production from early decays grows and becomes comparable to the contribution from late decays.  So $\psi$ is partially asymmetric with $Y_\psi^+ \simeq Y_{\rm FI}/2+\eta_D$ and $Y_{\psi}^- \simeq Y_{\rm FI}/2$. We take $R\ll1$, so that all $\chi$ decay into $\psi$, partially while being in thermal equilibrium (freeze-in) and partially at a later time (late decays). The $\psi$ abundance is $Y_{\psi}^+ + Y_\psi^-=Y_{\rm FI}+\eta_D$. On the other hand, the abundance of $S$ is determined by thermal freeze-out, once the symmetric population is annihilated away. Thus, $S$ freezes-out with an asymmetry $Y_{S}^+=\eta_D$. In this case the relative abundance between the two DM component and the total DM abundance are given by
	\begin{equation}
	\frac{\Omega_\psi}{\Omega_S}=\frac{m_\psi (\eta_D+Y_{\rm FI}^{})}{m_{S}\eta_D}\,,\qquad \frac{\Omega_{\rm DM}}{\Omega_B}\simeq 5=\frac{m_\psi (\eta_D+Y_{\rm FI})+\eta_D m_{S}}{\eta_B m_p}\,.
	\end{equation}
	
	\subsection*{Scenario 4}
	
	If $R\sim \mathcal{O}(1)$ for the values of Yukawa $y_\phi$ considered in the previous scenario (assuming that Eq.~\eqref{eq:LDS11} is satisfied), the population of $\psi,\bar{\psi}$ is produced partially by freeze-in and partially by late decays, whereas the decays into $S^\dagger$ washout the asymmetry in $S$, making it partially asymmetric. Therefore, in this scenario, the late decays determine the asymptotic nature of both DM components. The corresponding expressions for the DM abundance are
	\begin{align}\label{eq:DMgen4}
	\frac{\Omega_\psi}{\Omega_{S}}=\frac{m_{\psi}}{m_S} \frac{\eta_D+Y_{\rm FI}}{\eta_D(1+2R)}\,,\qquad \frac{\Omega_{\rm DM}}{\Omega_B}=\frac{m_\psi (\eta_D+Y_{\rm FI})+\eta_D (1+2R) m_{S}}{\eta_B (1+R) m_p}\,.
	\end{align}
	
	\subsection*{Mixed Scenario 3-4}
	
	Here we consider the same range of $y_\phi$ and $R$ as in the previous scenario but a larger $|y_S|$, which violates Eq.~\eqref{eq:LDS11}. The discussion for $\psi$ is the same as in Scenario 4. In analogy with the Mixed Scenario 1-2, $S^\dagger S$ annihilations erase the $S^\dagger$ from the thermal bath, leaving an asymmetric population of $S$ which survives the annihilations, so that
	\begin{align}
	\frac{\Omega_{\psi}}{\Omega_S}\simeq \frac{m_{\psi}}{m_S}\frac{Y_{\rm FI}+\eta_D}{\eta_D}\,,\qquad \frac{\Omega_{\rm DM}}{\Omega_B}\simeq 5=\frac{m_\psi(Y_{\rm FI}+\eta_D)+m_{S}\eta_D}{(1+R)m_p\eta_B}\,.
	\end{align}
	This scenario is a mixture between Scenarios 3 and 4.

	\subsection*{Scenario 5}
	
	For even larger Yukawas, in the range $2\times10^{-10}\sqrt{\eta_D/\eta_B}\lesssim y_\phi\lesssim 5 \times 10^{-7}$,
	the $\psi$ sector is mainly populated during freeze-in, while late decays only produce a sub-dominant component, i.e., $Y_{\rm FI} \gg \eta_D$. Therefore, the $\psi$ population is (almost) symmetric, $Y_\psi^+=Y_{\rm FI}/2+\eta_D\simeq Y_{\rm FI}$. Concurrently, $R$ is typically small because of the larger value of $y_\phi$ and therefore $S$ freezes with an asymmetric abundance, $Y_{S}^+=\eta_D$. 
	
	Therefore, DM is mostly symmetric in $\psi$ and asymmetric in $S$, produced by freeze-in and freeze-out, respectively. The relative abundance of the two species and the total DM abundance are
	\begin{equation}
	\frac{\Omega_{\psi}}{\Omega_{S}}=\frac{m_{\psi}}{m_S} \frac{Y_{\rm FI}}{\eta_D}\,,\qquad\frac{\Omega_{\rm DM}}{\Omega_B}\simeq 5 = \frac{m_\psi Y_{\rm FI}+m_{S}\eta_D}{m_p\eta_B}\,.
	\end{equation}
	
	\begin{figure}[!htb]
		\centering
		\includegraphics[width=0.6\textwidth]{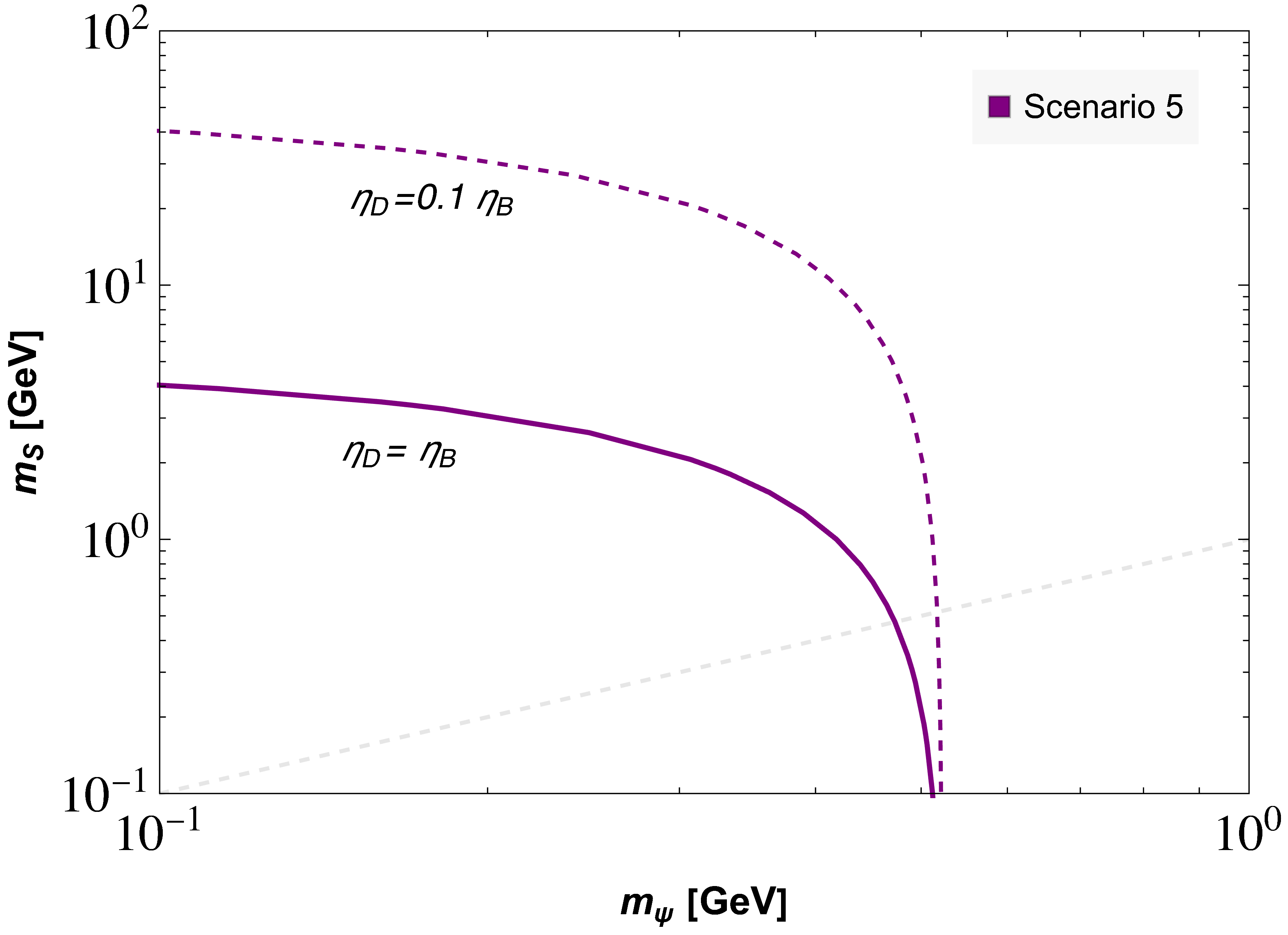}
		\caption{\label{fig:scenario3} Values of $m_\psi$ and $m_{S}$ for which the correct relic abundance can be reproduced for Scenario 5. The gray dashed line corresponds to $m_\psi = m_S$. We fix the parameters $|y_S|=2\times10^{-4}$ and $y_\phi=2 \times 10^{-10}$. With this choice, $R\simeq 5\times10^{-10}$ and $Y_{\rm FI}\simeq 10\eta_B$.}
	\end{figure}
	
	In Fig.~\ref{fig:scenario3} we show the mass ranges of $\psi$ and $S$ for which the DM relic abundance can be reproduced for different values of $\eta_D$. Notice that if $S$ is subdominant, the mass of $\psi$ is fixed, irrespective of the value of asymmetry. This is due to the fact that the DM mass is fixed by the freeze-in contribution in Eq.~\eqref{eq:FI}, which is (almost) independent of $m_\psi$. On the other hand, it depends on the mass of $\chi$. For the values used in the figure ($0.1\eta_B<\eta_D<\eta_B$, etc.), we see that DM could be mainly composed by light symmetric $\psi$ of mass around $500$ MeV, mainly by asymmetric GeV-ish $S$, or by a combination of them.	
	
As we have underlined above, for these values of $y_\phi$, $R$ is typically small. However, if we lower the $v_{B-L}$ scale it is possible to reach $R>1$ even in this scenario. We briefly discuss this possibility at the end of Section \ref{sec:ISS}.

	\subsection*{Scenario 6}
	
	Finally, when $R\gtrsim\mathcal{O}(10)$, for whatever value of $y_\phi$ compatible with it, the majority of the $\chi$ population decays into $S^\dagger$ after freeze-out of $S$ (but before BBN if $10^{-3}<|y_S|<0.1 (m_S/\text{GeV})$) and the asymmetry is completely washed-out. However, the populations of $S$ and $S^\dagger$ survive independently, as the rate of $S^\dagger S$ annihilations drops below the Hubble expansion rate by that time. Hence, the DM relic is almost completely made up by the symmetric population of $S$ and $S^\dagger$, i.e., $Y_{S}^+=Y_{S}^-$, while there is a negligible abundance of $\psi$ produced from early or late decays and leads to $m_{S}\simeq2.5\text{ GeV}\, \left({\eta_B}/{\eta_D} \right)\,$.	 The late restoration of the symmetric component would thus result in the enhancement of indirect detection signals.
	
Notice that for larger $|y_S|$ the decays take place while $S$ and $S^\dagger$ are still in equilibrium so that $S^\dagger S$ annihilations washout the asymmetry and one recovers the standard scenario of symmetric freeze-out in which the $S$ abundance is determined by the annihilation cross section $\sigma v_{S^\dagger S}$ instead of the asymmetry. Depending on $m_S$ and $\sigma v_{S^\dagger S}$ it may be possible to reproduce the correct relic abundance.  We did not consider this possibility in the main text.	

\section{Additional constraints}

\subsection{Decays of \texorpdfstring{$\varphi$}{varphi} to SM particles}\label{sec:varphidecay}

We ensure that $\varphi$ particles produced from $\chi$ decay fast enough into SM particles, so that the energy transfer from $\chi$ to SM radiation occurs before BBN. A rigorous bound arises from imposing $\tau_\chi+\tau_\varphi\lesssim 1$ sec. However, it is sufficient to check that at the time of $\chi$ decays, corresponding to $T=T_D$, the decays of $\varphi$ are fast compared to the Hubble rate and the bound in Eq.~\eqref{eq:BBN} applies (corresponding to $\tau_\chi\lesssim1$ sec.). The decay $\varphi\to$ SM can occur through the Higgs portal operator $\lambda_{H\phi}|\phi|^2|H|^2$ as both the scalars take a vev. Through this portal $\varphi$ decays into SM fermions, mainly (if kinematically allowed) into $\bar{b}b$ or $\bar{c}c$ or lighter species if $m_\varphi<$ GeV. The decay rate is  $\Gamma_\varphi=\sin^2\theta\times \Gamma_{h_\varphi\to \bar{f}f}$, where $\sin^2\theta$ is the mixing among $\varphi$ and the SM Higgs boson $h$ while $\Gamma_{h_\varphi\to \bar{f}f}\sim (m_\varphi/32\pi)(m_f/v_{\rm EW})^2$ is the decay width of a SM Higgs boson with mass $m_\varphi$ into SM fermions.

We require that the decay is fast at $T=T_D$ (when $\varphi$ are produced via $\chi$ decays), i.e., $\Gamma_\varphi/H|_{T=T_D}>1$. For $m_\chi\sim$ TeV, $m_\varphi\sim$ few GeV$>2m_c$ (or eventually $2m_b$) and $y_\phi\sim10^{-12}$ the ratio $\Gamma_\varphi/H$ at $T=T_D$ is $\gg 1$ even for small mixing angle, $\sin\theta\gtrsim10^{-8}$.
Even lighter $m_\varphi\gtrsim$ MeV is allowed as $\varphi$ can decay to $\bar{u}u,\bar{d}d,\bar{e}e$, with a larger mixing angle $\sin\theta>5\times 10^{-4}$. Bounds from LEP constrain the mixing of a GeV-ish scalar to the SM Higgs to be $\sin\theta<0.1$ \cite{Clarke:2013aya}. Therefore, the condition of Eq.~\eqref{eq:BBN} is valid. 
$m_\varphi$ lighter than MeV cannot decay to any SM fermion. However it could decay into 2 photons through the effective Higgs-photon interactions. We do not study this possibility and we consider $m_\varphi>$ MeV. This also implies $m_{S}\gtrsim$ MeV.

\subsection{Gauge boson spectrum and constraints on massless $A'_\mu$} \label{sec:masslessA}

The gauge boson masses after symmetry breaking are given by 
\begin{equation}
\begin{cases}
m_{A'}^2=0\,, \\
m_{Z_{D}}^2=2g_D^2v_\phi^2\left(1 + \mathcal{O}\left(g_X^2,\left({v_\phi}/{v_{B-L}}\right)^2,\kappa^2\right) \right)\,,\\
m_{Z_{B-L}}^2=8g_{B-L}^2v_{B-L}^2 \left(1+ \mathcal{O}\left(g_X^2,\left({v_\phi}/{v_{B-L}}\right)^2,\kappa^2\right)\right)\,,
\end{cases}
\end{equation} 
where we used that $g_X\ll1$, $\kappa\ll1$ and $v_\phi\ll v_{B-L}$.
The mixing among them at the leading order in the small expansion parameters (omitting Lorentz indices) can be expressed as
\begin{equation}{\label{eq:gaugekm}}
\begin{cases}
A^{'0}\simeq A^{'}-g_X/g_D \, Z_{D}\,, \\
Z^0_D\simeq Z_D+g_X/g_D \,A^{'}-\left[(g_D/4g_{B-L})(v_\phi^2/v_{B-L}^2)+\kappa\right]Z_{B-L}\,, \\
Z_{B-L}^0\simeq Z_{B-L}+(g_D/4g_{B-L})(v_\phi^2/v_{B-L}^2)(1+\kappa g_{B-L}/g_D)Z_{D}^{'}
-(g_X/4g_{B-L})(v_\phi^2/v_{B-L}^2) A' \nonumber\,. 
\end{cases}
\end{equation}

The massless gauge bosons $A'_\mu$ only interact with the fermion $\psi_0$.
Taking into account the fermion mixing, the Lagrangian contains the interaction terms $\bar{\psi}\psi A'$ (suppressed by $g_X$), $\bar{\psi}\chi A'+{\rm H.c.}$ (suppressed by $g_X\epsilon_f$) and $\bar{\chi}\chi A'$ (suppressed by $g_X\epsilon_f^2$). These vertices give rise to scattering processes as $\bar{\chi}\chi\to A'A',\bar{\psi}\psi\to A'A',...$ or decays $\chi\to A'\psi$. 
However, given the smallness of $g_X$ (and $\epsilon_f$) these processes are extremely suppressed and no sizeable population of $A'$ (which in principle could contribute to dark radiation) is produced.

We can give an upper bound on the value of the gauge coupling coming from long-range force experiments. Indeed the mixing between $A'$ and $Z_{B-L}$ induces an interaction $g_{\rm eff}A'_\mu J^\mu_{B-L}$ with $g_{\rm eff}\simeq (g_X/g_{B-L})(v_\phi/v_{B-L})^2$. For the reference values of this work, $v_\phi\sim$ TeV and $v_{B-L}\sim10^{11}$ GeV, this corresponds to an effective coupling $g_{\rm eff}\sim 10^{-16}g_X/g_{B-L}$. Long-range force experiments constrain the coupling to the $B-L$ current to be $g_{\rm eff}\lesssim 10^{-24}$ \cite{Heeck:2014zfa}, which implies $g_X\lesssim 10^{-8}$ (for $g_{B-L}\sim \mathcal{O}(1)$). This is comparable with the condition for $\psi$ not to reach thermal equilibrium.

\subsection{Implications of fermion mixing}\label{sec:fermionmix}

Diagonalization of the fermionic sector at leading order in $\epsilon_f$ leads to the following masses for the mass-eigenstates $\chi$ and $\psi$,
\begin{equation}
\begin{cases}
m_\psi=m^0_\psi-\epsilon_f^2(m^0_\chi-m^0_\psi)+\mathcal{O}(\epsilon_f^3) \,,\\
m_\chi=m^0_\chi+\epsilon_f^2(m^0_\chi-m^0_\psi)+\mathcal{O}(\epsilon_f^3) \,.
\end{cases}
\end{equation}
The fields in the mass basis are related to the original fields as
\begin{equation}
\begin{cases}
\psi=(1-\epsilon_f^2/2)\,\psi_0-\epsilon_f\,\chi_0+\mathcal{O}(\epsilon_f^3)\,, \\
\chi=(1-\epsilon_f^2/2)\,\chi_0+\epsilon_f\,\psi_0+\mathcal{O}(\epsilon_f^3)\,.
\end{cases}\,
\end{equation}
Fermion mixing induces new interactions due to the $\bar{\psi}Z_D\chi$ coupling, suppressed by $\epsilon_f$, leading to new production processes such as $\bar{\chi}\chi\to \bar{\chi}\psi$. However, these scattering processes are typically sub-dominant to decays due to suppression by $y_\phi$ and $\epsilon_f$ as well as a phase space suppression (taking $\Delta =1$), and thus can be neglected.

In addition to the above, the mixing could also lead to an additional contribution to $\psi$ production. For $T>v_\phi$, the particles in the thermal bath are $\chi_0$ and $\psi_0$ and there is no mixing. For $T<v_\phi$, the vev of $\phi$ induces the mixing so that the states $\chi_0$ contain a small $\psi$ component, proportional to $\epsilon_f$. Therefore, $\chi_0\to\psi$ conversions contribute to the final $\psi$ abundance (this is analogous to the production of sterile neutrino from active neutrino mixing). This is also freeze-in process as the population of $\psi$ is produced non-thermally from a small coupling (the fermion mixing). In analogy with sterile neutrino DM, the interaction rate of $\psi$ is $\Gamma_\psi\simeq\epsilon_f^2\Gamma_\chi$, with $\Gamma_\chi=n_\chi\sigma_{\bar{\chi}\chi}$. If $v_\phi\lesssim m_\chi$ the $\chi$ particles are non-relativistic but still in equilibrium when the mixing is generated. So, $n_\chi=n_\chi^{\rm eq}$ and these processes contribute to the symmetric component of $\psi$. Using the number density at equilibrium we checked that for the typical values of our parameters ($m_\chi\simeq3.5$ TeV, $g_D\simeq 0.5$, $v_\phi\sim$ TeV) this contribution is at most comparable (but not larger) than the one from decays. A detailed study of this contribution, solving the full Boltzmann Equations (or using the density matrix formalism) is beyond the scope of this work. Therefore, we neglect this contribution, having in mind that it would change the total (symmetric) freeze-in contribution by at most an $\mathcal{O}(1)$ factor.

\section{Contributions to operator \texorpdfstring{$\mathcal{O}_6=\bar{L}\tilde{H}S\phi^\dagger \psi$}{O6}}{\label{sec:stability}}

In Section~\ref{sec:pheno}, we discussed the stability of the two DM components $S$ and $\psi$ and the possible observation of a neutrino line from the decay of one component into the other. As we saw, the decay is mediated by the dimension-6 operator 
\begin{equation}
\mathcal{O}_6=\bar{L}\tilde{H}S\phi^\dagger \psi,
\end{equation}
which is generated at low energy by first integrating the right-handed neutrino field and then the fermion $\chi$, as discussed in the main text. A second contribution to $\mathcal{O}_6$ may arise if we assume that the theory contains the dimension-$5$ operator
\begin{equation}
\mathcal{O}_5=\frac{\bar{N}_R (S\phi^\dagger) \psi}{\Lambda_{\rm UV}}\,,
\end{equation}
where $\Lambda_{\rm UV}$ is a cut-off scale parametrising the UV completion of our model at scales above $v_{B-L}$. Integrating out $N_R$ gives rise to
\begin{equation}
\frac{y_\nu}{\Lambda_{\rm UV}M_{N_1}}\mathcal{O}_6\,.
\end{equation}
The condition on the lifetime $\tau_S^{\rm re-sc}>10^{23}$ s (taking $m_S>m_\psi$), is now satisfied if
\begin{equation}
\Lambda_{\rm UV}\gtrsim 10^{16}\text{GeV}\left(\frac{m_\nu}{0.05\text{eV}}\right)^{1/2}\left(\frac{10^{11}\text{GeV}}{M_{N_1}}\right)^{1/2}\left(\frac{v_\phi}{\text{TeV}}\right)\left(\frac{m_S}{50\text{GeV}}\right)^{1/2}\,.
\end{equation}
This scale must be at most Planckian, i.e., $\Lambda_{\rm UV}\lesssim M_{\rm Pl}$, which implies a bound on the mass of the heaviest of the 2  DM particles
\begin{equation}\label{meq:mSlimit2}
m_S\lesssim 10^4\text{ TeV}\left(\frac{0.05\text{eV}}{m_\nu}\right)\left(\frac{M_{N_1}}{10^{11}\text{GeV}}\right)\left(\frac{700\text{GeV}}{v_\phi}\right)^2\,.
\end{equation}
Finally, $\mathcal{O}_6$ could be directly generated by UV physics as ${\mathcal{O}_6}/{\Lambda_{\rm UV}^{'2}}$. In this case we just need 
\begin{equation}
\Lambda'_{\rm UV}\gtrsim 10^{15}\text{ GeV}\left(\frac{v_\phi}{\text{TeV}}\right)^{1/2}\left(\frac{m_S}{50\text{GeV}}\right)^{1/4}\,,
\end{equation}
which gives a weaker constraint. Notice that the same bounds apply on $m_\psi$ if $m_\psi>m_S$. Summarising, Eqs.~\eqref{meq:mSlimit1} and \eqref{meq:mSlimit2} set an upper bound on the mass of the heavier between $\psi$ and $S$. Therefore, this shows that it is quite natural that both $\psi$ and $S$ are stable on cosmological scales, and that they both contribute to the DM relic abundance and respect current limits from neutrinos.

\bibliographystyle{JHEP}
\bibliography{JHEPFINAL}

\end{document}